\documentclass[usenatbib]{mnras}
\usepackage{amsbsy}
\usepackage{amsmath}
\usepackage{amssymb}

\usepackage{amsfonts}
\usepackage{graphicx}
\usepackage{color}
\usepackage{subfig}

\makeatletter
\newcommand\footnoteref[1]{\protected@xdef\@thefnmark{\ref{#1}}\@footnotemark}
\makeatother

\title[Modelling the UV spectrum of SDSS-III/BOSS galaxies]{Modelling the UV spectrum of SDSS-III/BOSS galaxies: hints towards the detection of the UV upturn at high-z}
\author[Claire Le Cras]{Claire Le Cras$^{1}$\thanks{E-mail: claire.lecras@port.ac.uk}, Claudia Maraston$^{1}$\thanks{E-mail: claudia.maraston@port.ac.uk}, Daniel Thomas$^{1}$\thanks{E-mail: daniel.thomas@port.ac.uk}, and Donald G. York$^{2}$\\
$^{1}$ICG, University of Portsmouth, Dennis Sciama Building, Burnaby Road, Portsmouth, PO1 3FX, UK\\
$^{2}$Dept. of Astronomy and Astrophysics and The Fermi Institute, The University of Chicago, 5640 South Ellis Avenue, Chicago, IL 60615, USA}

\begin{document}

\maketitle

\label{first page}

\begin{abstract}
We exploit stellar population models of absorption line indices in the ultraviolet (from 2000 - 3200\AA) to study the spectra of massive galaxies. Our central aim is to investigate the occurrence at high-redshift of the UV upturn, i.e. the increased UV emission due to old stars observed in massive galaxies and spiral bulges in the local Universe. We use a large ($\sim 275,000$) sample of $z \sim$ 0.6 massive (M$^{*}$/M$_{\odot}>$ 11.5) galaxies using both individual spectra and stacks and employ a suite of models including a UV contribution from old populations, spanning various effective temperatures, fuel consumptions and metallicities. We find that a subset of our indices; Mg I, Fe I, and BL3096, are able to differentiate between old and young UV ages. We find evidence for old stars contributing to the UV in massive galaxies, rather than star formation. The data favour models with low/medium upturn temperatures (10,000 - 25,000K) consistent with local galaxies, depending on the assumed metallicity, and with a larger fuel \textbf{($f \sim 6.5\cdot10^{-2}~\textnormal{M}_{\odot}$)}. Models with one typical temperature are favoured over models with a temperature range, which would be typical of an extended horizontal branch. 
Old UV-bright populations are found in the whole galaxy sample ($92\%$), with a mass fraction peaking around 20-30\%. Upturn galaxies are massive and have redder colours, in agreement with findings in the local Universe. We find that the upturn phenomenon appears at $z \sim 1$ and its frequency increases towards lower redshift, as expected by stellar evolution of low mass stars. Our findings will help constrain stellar evolution in the exotic UV upturn phase. 
\end{abstract}


\section{Introduction}

One of the most effective ways to learn about galaxies is to study their spectral energy distribution (SED) which contains a wealth of information about both their gas and stellar populations. Absorption lines and the stellar continuum component of a spectrum can be used to derive ages, element abundances, star formation histories, and formation epochs (e.g. \citet{Worthey:1992}, \citet{GonzalezDelgado:1999}, \citet{Trager:2000}, \citet{Kauffmann:2003}, \citet{Rose:2005}, \citet{Thomas:2005, Thomas:2010, Thomas:2011}, \citet{Tojeiro:2013}, \citet{Wilkinson:2015}) whereas emission lines can be used to discern information regarding ongoing star formation activity, gas kinematics, and black hole accretion (e.g. \citet{Pettini:2001}, \citet{Kauffmann:2003b}, \citet{Shapley:2004}, \citet{Tremonti:2004}, \citet{Sarzi:2006}, \citet{Schawinski:2007}).

In such analyses, properties are derived by comparing data with model spectra where we are confronted with several degeneracies most noticeably: age, metallicity, and dust. Element abundance ratios help break these degeneracies (\citet{Thomas:2005}) as does the use of a large wavelength range (\citet{Pforr:2012}). On the other hand, very often only a small spectral window is available in data. 

In this paper we focus on the UV spectrum, which is traditionally used to gain clues on star formation. 

The UV traces the hot component of stellar populations. In younger galaxies this is composed of luminous O and B-type stars. In old populations old stars may become UV-bright after sufficient mass loss, as shown in phenomena such as the UV upturn in local elliptical galaxies (e.g. \citet{Dorman:1995}, \citet{Burstein:1988}, \citet{GreggioRenzini:1990}, \citet{O'Connell:1999}, \citet{Brown:2000}, \citet{Yi:2005}, \citet{Yi:2008}) and the extreme horizontal branch in globular clusters (e.g. \citet{deBoer:1985}, \citet{Ferraro:1998}, \citet{Lee:2005}, \citet{Rey:2009}).

The UV upturn is seen as an increase in the UV flux of galaxy spectra short-ward of 2500\AA. The leading hypothesis on its origin suggests that this increase in flux is due to old, hot, low-mass stars (\citet{GreggioRenzini:1990}). These may be stars evolving on the extreme horizontal branch and their progeny (either metal poor or metal rich, with possible element enhancement, including helium). Other possible sources include binary star systems (\citet{Han:2007}) and post-asymptotic giant branch (PAGB) stars in the planetary nebula stage. Contributions from young populations can usually be excluded based on other grounds, e.g. strong C IV lines not observed in old galaxies, and  the overall shape of the UV spectrum). We direct the reader to \citet{O'Connell:1999}, \citet{Brown:2004}, and \citet{Yi:2008} for reviews of these hypotheses and issues.

The interest in understanding the origin of the UV upturn is twofold. Firstly, it is a phenomenon related to exotic stellar evolution which still eludes a first-principle understanding. Secondly, in absence of an alternative explanation, a UV enhancement is usually interpreted in terms of star formation. Hence, the general question on what causes the UV light emission in galaxies impacts on our understanding of galaxy formation.

As a general finding in the local Universe, the spectral slope seen in the UV upturn remains roughly constant and is found to be dependant on bulk galaxy properties. \citet{Burstein:1988} investigated the dependence of the upturn on the stellar population and dynamical properties of a large sample of early-type galaxies. They found a clear non-linear correlation of the UV-optical colours with the strength of the Mg$_2$ absorption line index, with bluer UV galaxies having strong line strengths. The correlations of \citet{Burstein:1988} were weakly recovered by \citet{Boselli:2005} for early-type members of the Virgo cluster, however it is unclear how these correlations extend to low-luminosity systems. \citet{Bureau:2011} recover the negative correlation between FUV - V colour and Mg$_2$ line strength using 48 nearby early-type galaxies in the SAURON sample, suggesting a positive dependence of the UV upturn on metallicity.  

On the modelling side, \citet{Maraston:2000} show how the observed UV upturn of the Burstein sample can be fit using models containing an old component at high-metallicity and plausible fuel consumptions. Here we shall use the same modelling, also varying the two key parameters, effective temperature and fuel consumption.

As mentioned before, the question is whether one can distinguish the contribution to the UV from old and young stars. \citet{Yi:2005} have shown that a large fraction of galaxies (over 30\%) are found to have remarkably strong NUV and FUV emission using large samples of early type galaxies from the Sloan Digital Sky Survey. This UV emission is far more powerful than the emission found in traditional upturn galaxies such as NGC 1399 and 4552 and deemed too powerful to be accounted for by any current model of the upturn. They concluded that the majority of these UV bright early type galaxies were inconsistent with a classic upturn and instead experienced residual star formation, with only a small fraction of elliptical galaxies showing a strong upturn (generally limited to the brightest cluster galaxies). Further to this, using a classification scheme based on UV broadband colours, \citet{Yi:2011} found that only 5\% of cluster elliptical galaxies show a UV upturn, with the other 27\% and 68\% of their working sample (built from the \citet{Yoon:2008} galaxy cluster catalog) being classified as "recent star formation" and "UV-weak" respectively. This hypothesis was elaborated on by \citet{Schawinski:2007b} and \citet{Kaviraj:2007} who concluded that residual star formation of a few per cent of the galaxy mass is common in early type galaxies with low stellar mass  ($10^{10}~\textnormal{M}_{\odot}$). \citet{Thomas:2010} reinforce that residual star formation in low-mass galaxies is also correlated to environment. 

One question that has not been addressed in detail so far is the redshift evolution of the UV upturn. If the upturn has indeed an origin in the evolved stellar component of a galaxy, it should appear at a relatively modest redshift, when stars in galaxies are sufficiently old. This evolution was first investigated in \citet{Brown:1998, Brown:2000, Brown:2003} via far-UV images of 4 - 8 elliptical galaxies at $z$ = 0.375, 0.55, and 0.33 respectively. They find the strength of the upturn to show little evolution between $z$ = 0.33 and $z$ = 0.55 with no galaxies being as blue as local, $z$ = 0,  strong upturn galaxies. Their results do not support residual star formation as the source of the UV flux seen in these galaxies as this would lead to a stronger UV flux at $z$ = 0.55, not weaker as shown by the data.

The Brown sample includes only a few objects and is based on photometry. The further study of the redshift evolution of the UV upturn, up to $z \sim$ 1, is one of the aims of this work, for which we use a very large sample of galaxy spectra (see below). In addition, the UV upturn is usually analysed using broadband photometry, while here we shall make the first attempt to use UV spectral indices rather than colours.

We use stellar population models of UV indices from \citet{Maraston:2009}, hereafter M09, who computed both empirically based as well as fully theoretical models. The far-UV extension of these models was tested in M09 using young star clusters in the Magellanic Clouds with independently known ages and metallicities, and a set of 11 indices blueward of 2000\AA \ were found to be able to recover their ages and metallicities well. This test could not be performed in the mid-UV due to the lack of data. The M09 models were also used in \citet{McDonald:2014} to probe the young stellar populations within a cooling filament in Abell 1795, using deep far-UV spectroscopy obtained using the \textit{Cosmic Origins Spectrograph} on HST. 

This paper also uses an extension of the M09 models which includes several models containing a UV upturn contribution. We explore the effect of such populations on the UV indices to determine their usefulness in identifying old UV-bright populations, such as those that might contribute to the UV upturn, and their contribution by mass. We also investigate the redshift evolution of these populations in an effort to constrain the onset of the UV upturn.

We apply these models to galaxies from the Sloan Digital Sky Survey (SDSS) - III Baryon Oscillation Spectroscopic Survey (BOSS, \citet{Dawson:2013}). BOSS was created to detect the characteristic scale imprinted by baryon acoustic oscillations in the early universe using as probes the spatial distribution of luminous and massive galaxies and quasars. Hence in addition to the cosmological experiment the survey provides an outstanding spectroscopic sample of massive galaxies ideally suited to probing the formation and evolution of galaxies in the Universe. 

The BOSS sample is ideal for exploring old UV-bright stellar populations and their appearance and evolution as a function of cosmic time. BOSS is mainly composed of massive, red galaxies ($\textnormal{M}>10^{11} \textnormal{M}_{\odot}$) distributed to $z \sim$ 1, with an average $z \sim$ 0.57. BOSS galaxies are thereby the most likely progenitors of present day, massive galaxies displaying the UV upturn. Their redshift distribution allows us to have a view on the past few billion years, which is ideal to explore the onset period of low mass stars as generators of UV light in quiescent galaxies. The wavelength coverage of the BOSS spectra (3600 - 10,400\AA) allows us to access mid-UV spectral indices above $z$ = 0.6.  We use Data Release 12 (DR12) which contains spectra from all previous data releases of BOSS as well as all imaging and spectra from previous SDSS data releases.

The structure of this paper is as follows. Section \ref{sec:mod} summarises the models used in this work and recalls the adopted line index system. In Section \ref{sec:data} we introduce the data used, taken from SDSS - III / BOSS, and discuss the coverage of our models in comparison to the data in Section \ref{sec:cover}. In Section \ref{sec:analysis} we present our analysis of the UV models as applied to galaxies. Section \ref{model:ex} explores the properties of galaxies found to have old UV-bright populations in our working sample and Section \ref{sec:local} looks for high $z$ analogues of local upturn galaxies. In Section \ref{sec:conc} we provide a summary and conclusions.

\section{Models of UV Line Indices}
\label{sec:mod}

The models used here are taken from \citet{Maraston:2009} or newly computed for this work (see Section \ref{model:th}). 

M09 model UV narrowband indices which were defined by \citet{Fanelli:1992}, hereafter F92, using data from the \textit{International Ultraviolet Explorer} (IUE). Indices were determined by F92 to characterise the continuum and spectral lines of the IUE stellar spectra in two spectral regions; 1230 - 1930\AA \ and 1950 - 3200\AA. The first set of indices, defined in the far-UV, were created for the analysis of active star-forming galaxies whereas the second set, in the mid-UV, were constructed for the analysis of the properties of spectral features prominent in old populations. The F92 indices mostly trace lines of Fe and Mg, and blends of different species. M09 utilise the full line index system defined by F92, see their Table 3, which is comprised of 11 indices in the far-UV (1270 - 1915\AA) and 8 indices in the mid-UV (2285 - 3130\AA). Synthetic line indices were obtained for simple stellar populations (SSP), i.e. chemically homogeneous and instantaneous bursts, for several metallicites ($2 Z_{\odot}, Z_{\odot}, \frac{1}{2}Z_{\odot}$ and $\frac{1}{10}Z_{\odot}$), and ages (1 Myr $\le t \le$ 1 Gyr or above depending on the available stellar data, see Section \ref{model:em}). 

The M09 models, based on the (\citet{Maraston:1998, Maraston:2005}, hereafter M05) evolutionary population synthesis (EPS) code are computed using a twofold strategy, producing both \textit{empirically based models as well as theoretical ones}, as we describe below. 

\subsection{Semi-empirical Model Indices}
\label{model:em}

The M09 empirically-based models use the index description as a function of stellar parameters as measured on the real IUE stellar spectra collected by F92. The data consists of low-resolution (6\AA) observations of 218 stars in the solar neighbourhood, covering 1150 - 3200\AA \  and are compiled in the "IUE Ultraviolet spectral atlas" (\citet{Wu:1983, Wu:1991}). The library consists of 56 stellar groups, classified by spectral type and luminosity class. Out of these the 47 groups classified as having solar metallicity were selected as input for the construction of {\it fitting functions} (FF), which are then plugged into the M05 EPS code. FFs are analytical approximations that describe the line indices as functions of stellar parameters (T$_{\textnormal{eff}}$, log $g$, [$Z$/H]). The FF method is a convenient approach used by many authors to include an empirical calibration in a population synthesis code. An advantage of this method is that fluctuations in the spectra of stars with similar atmospheric parameters become averaged out.

As the observed stars have mostly solar chemical composition the FF were initially built to depend only on T$_{\textnormal{eff}}$ and log $g$. Metallicity effects were incorporated by means of the high resolution synthetic Kurucz spectra calculated by \citet{RodriguezMerino:2005}, see Section \ref{model:th1}. Model spectra with log $g = 4$ were considered, as the most important contributors to the UV light are stars at the main sequence turnoff whose gravity does not vary appreciably.

To estimate fractional metallicity corrections the separate effects of abundance changes were considered on both the absorption feature and the pseudo-continuum fluxes. The FF calculated in this way were plugged into the M05 EPS code to produce integrated indices for metallicities 2, $\frac{1}{2}$ and $\frac{1}{10}Z_{\odot}$.

The FF were calculated to be suitable for population ages from 1 Myr up to 1 Gyr. This is due to the range of stars observed by the IUE and the fact that the UV flux usually drops off after this age. Extrapolating the FF out to older ages would cause the EWs to continue to increase which we show not to be the case for several indices using theoretical models, see Figure \ref{fig:ex_trend}.

\subsection{Theoretical Model Indices}
\label{model:th1}

In parallel to the FF-based approach of the empirical model, integrated line indices were calculated via direct integration on the synthetic SEDs of stellar population models computed using the high resolution theoretical library UVBLUE by \citet{RodriguezMerino:2005} as input to the M05 EPS code. UVBLUE is based on LTE calculations executed with the ATLAS9 and SYNTHE codes developed by \citet{Kurucz:1979}. 
Many of these indices were studied in theoretical stellar spectra by \citet{Chavez:2007} who found the synthetic indices to follow the general trends depicted by those computed from empirical databases by exploring their behaviour in terms of leading stellar parameters (T$_{eff}$, log $g$).
Spectra are provided for a wide range of chemical compositions, temperatures, and gravities with a wavelength range of 850 - 4700\AA, and high spectral resolution ($\frac{\lambda}{\Delta \lambda}\sim 10,000$)\footnote{Both the full high-resolution and further downgraded versions are available at \it{www.maraston.eu}}. 

For this work the calculations of UV indices on the M09 theoretical model has been extended to also cover old ages (up to 14 Gyr).

\subsubsection{UV-upturn Models}
\label{model:th}

Here we use extensions of the theoretical model produced by C. Maraston which incorporate old hot stellar population components. Two sets of models were calculated, which differ by the source of the UV light. 

\setcounter{figure}{0}
\begin{figure*}
\centering
\subfloat[The mid-UV region of $Z_{\odot}$ upturn models]{%
\includegraphics[width=0.85\textwidth]{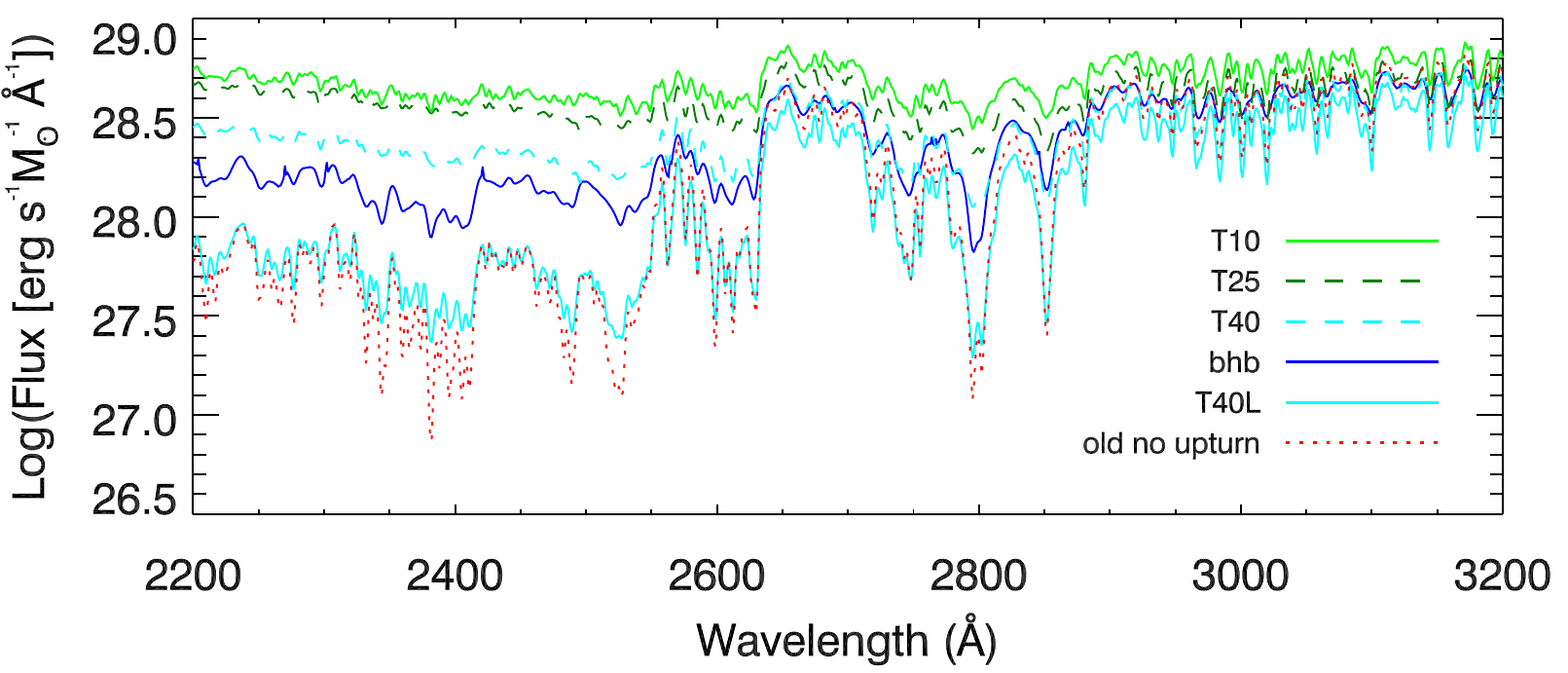}}\\
\subfloat[The mid-UV region of $2Z_{\odot}$ upturn models]{%
\includegraphics[width=0.85\textwidth]{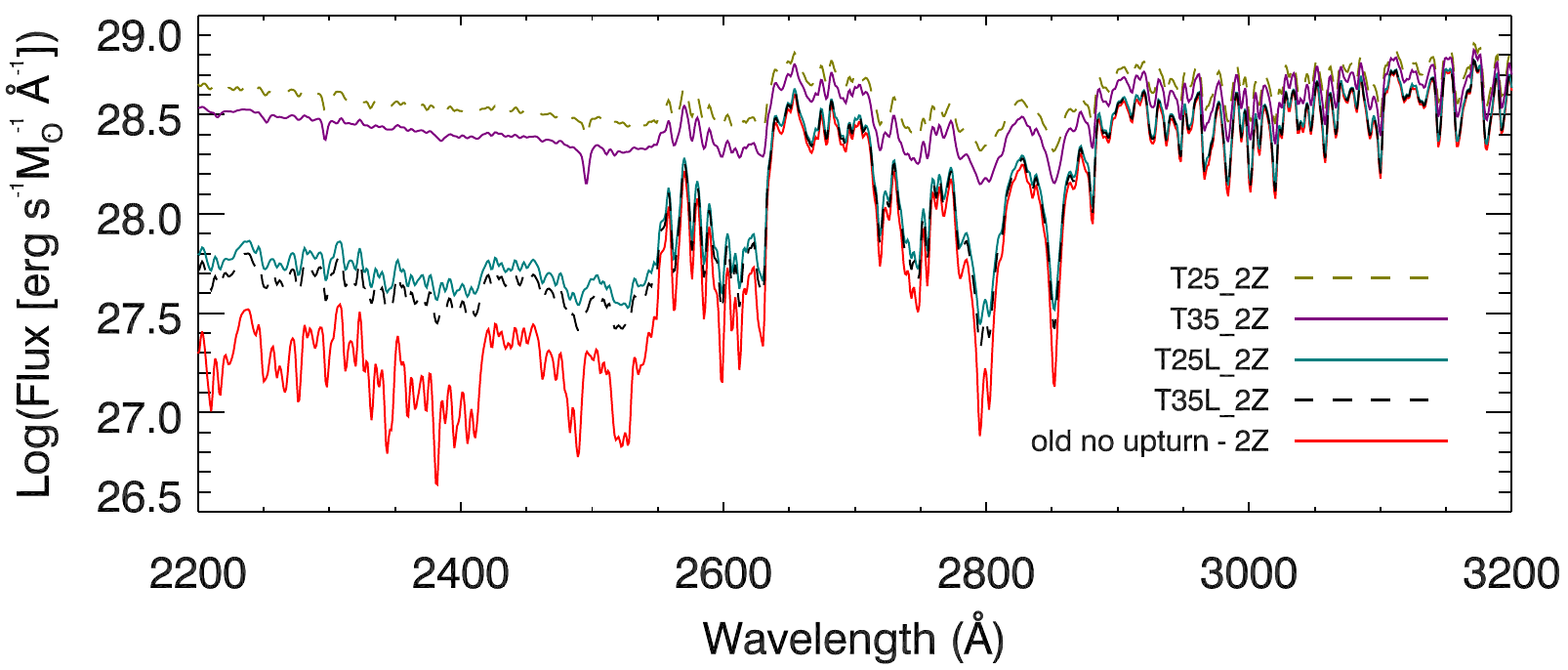}}
\caption{The mid-UV region of smoothed theoretical spectra showing multiple upturn models. Each spectrum shows a 5 Gyr, solar metallicity population. (a) The $Z_{\odot}$ models. The solid light green and dashed dark green lines show the T10 and T25 models with the solid blue line showing the bhb model. The dashed cyan line shows the T40 model with the solid cyan line showing the T40L model with lower fuel. The dotted red line shows the old model without an upturn component. (b) $2Z_{\odot}$ models. The dashed olive and solid turquoise lines show the T25\_2Z and T25L\_2Z models respectively with the solid purple and dashed black lines the T35\_2Z and T35L\_2Z models. The solid red line shows a model without an upturn component.}
\label{fig:upturns}
\end{figure*}

\setcounter{table}{0}
\begin{table}
\caption{A list of theoretical model names and parameters for the PAGB models used in this 5, ordered as increasing temperature and fuel.}
\begin{tabular}{l || c | c | c | l}
Name & Metallicity & Temperature (T$_{\rm eff}~/K$) & Fuel ($f$)$/\textnormal{M}_{\odot}$\\ \hline
\hline
T10 & $Z_{\odot}$ & 10,000 & $6.5\cdot10^{-2}$ \\
T25 & $Z_{\odot}$& 25,000 & $6.5\cdot10^{-2}$ \\
T25L\_2Z & $2Z_{\odot}$ & 25,000 & $6.5\cdot10^{-3}$ \\
T25\_2Z & $2Z_{\odot}$ & 25,000 & $6.5\cdot10^{-2}$ \\
T35L\_2Z & $2Z_{\odot}$ & 35,000 & $6.5\cdot10^{-3}$ \\
T35\_2Z & $2Z_{\odot}$ & 35,000 & $6.5\cdot10^{-2}$ \\
T40L & $Z_{\odot}$ & 40,000 & $6.5\cdot10^{-3}$ \\
T40 & $Z_{\odot}$ & 40,000 & $6.5\cdot10^{-2}$ \\
\label{tab:models}
\end{tabular}
\end{table}

One type of model has a blue horizontal branch (BHB) at high-metallicity as described in M05. This is calculated by applying a Reimers' mass loss with efficiency parameter $\eta=0.55$\footnote{The canonical value calibrated with Milky Way globular clusters and used throughout the models is $\eta=0.33$} to the post-main sequence stellar track corresponding to a given age. As a result of the increased mass-loss, the phase of helium burning contains sub-phases spent at progressively higher temperatures, which raises the emission at short wavelength. The range of temperatures spanned depends on the assumed mass-loss (see M05 for details). This first model, referred to throughout the paper as bhb, has a {\it continuous} temperature distribution, with fuel consumption calculated at each temperature location as proper to the corresponding stellar track (see M05).

Studies of the dwarf spheroidal galaxy M32 (\citet{Brown:2000b} and \citet{Brown:2008}) have shown the presence of hot horizontal branch stars in its core with a lack of UV-bright PAGB stars. By constructing an ultraviolet colour-magnitude diagram of the hot HB population they find that only $\sim$2\% of the HB population is hot enough to produce significant UV emission. However, this is sufficient to provide the vast majority of the emission seen in the galaxy, implying PAGB stars are not a significant source of emission. It should be noted that M32 is a dwarf spheroidal galaxy whereas the galaxies investigated in this paper are the most massive ellipticals therefore the upturn mechanism may differ between the two populations.

\setcounter{figure}{1}
\begin{figure*}
\centering
\subfloat[$Z_{\odot}$ upturn models]{%
\includegraphics[width=0.8\textwidth]{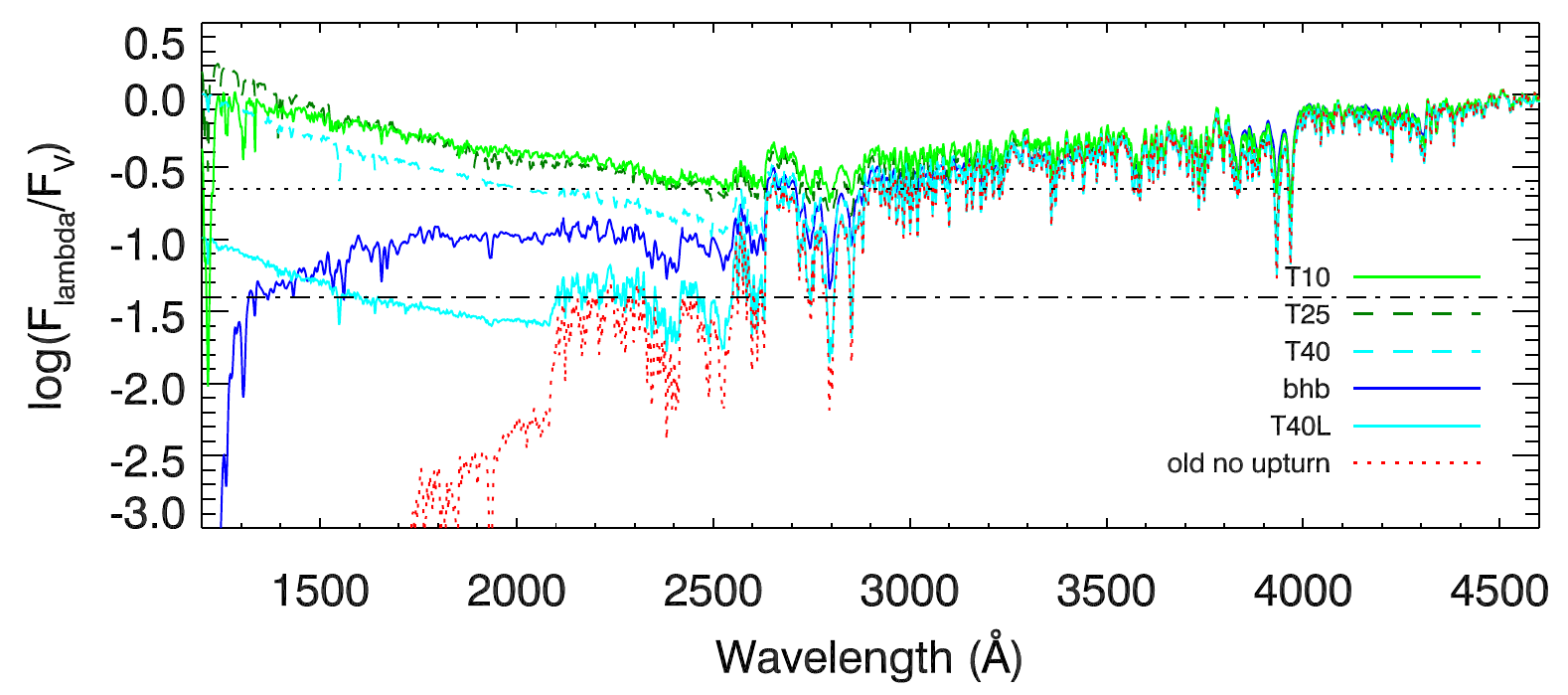}}\\
\subfloat[$2Z_{\odot}$ upturn models]{%
\includegraphics[width=0.8\textwidth]{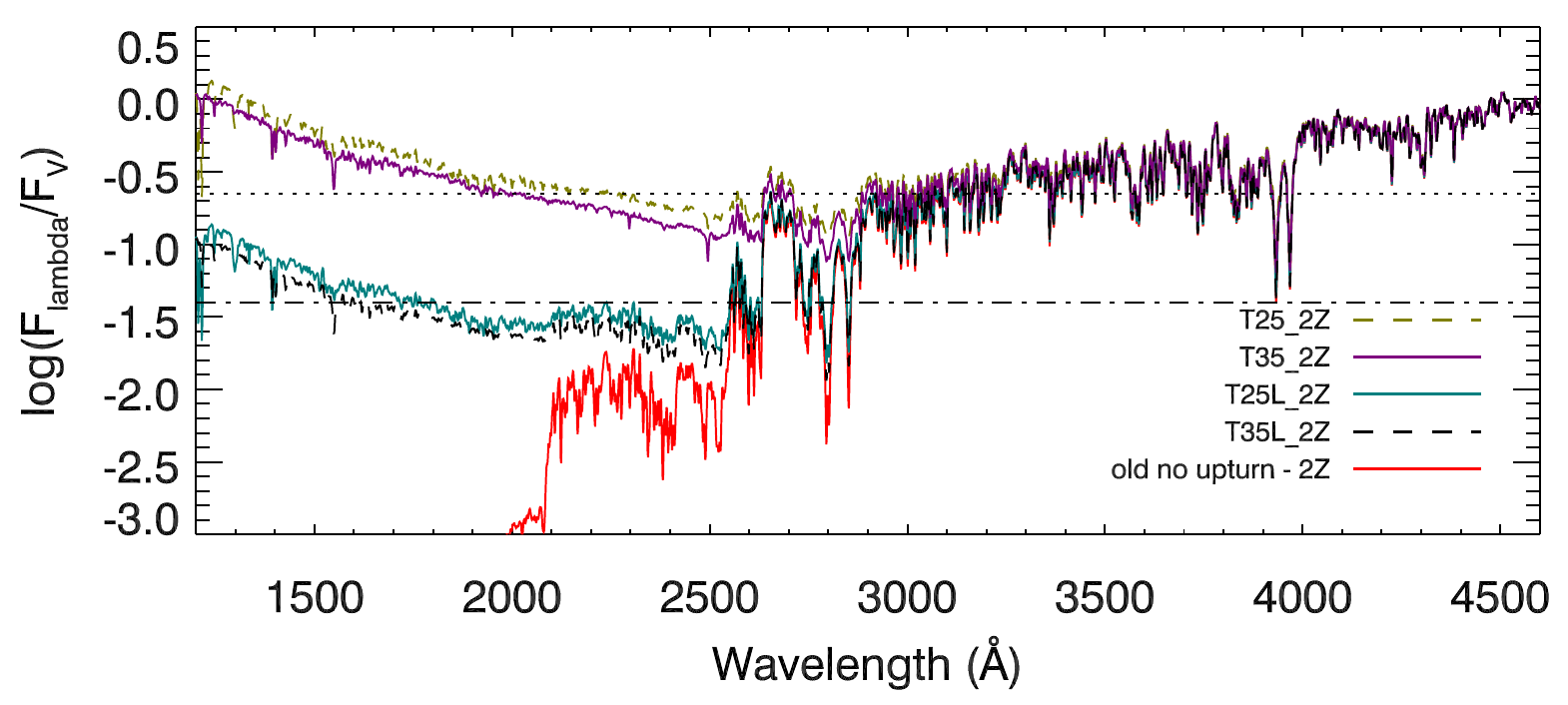}}
\caption{Spectral energy distributions of 5 Gyr upturn models as a function of wavelength, normalised to 4699\AA. The dash-dot and dotted lines show the y-intercept of local galaxies featuring an upturn (NGC3379 and NGC4649 respectively), taken from Figure 5 from \citet{Maraston:2000}}
\label{fig:upturn_limits}
\end{figure*}

The other type of model assumes that the hot phase in old populations is localised at certain well-defined temperatures, aiming at simulating a PAGB evolution. As a starting value we consider a model with a temperature T$_{\rm eff}=25,000$~K, which was found by \citet{Maraston:2000}, hereafter MT00, to reproduce the shape of the average UV upturn of four local objects (three massive early-type galaxies and the bulge of M31). 

However, because the UV upturn strength varies among galaxies and may also vary with redshift in a non trivial way, we further explore other combinations of temperatures and also fuels, see Table \ref{tab:models} for details of the combinations and naming convention.

The numerical value is the fuel consumption appropriate to an intermediate/old helium burning phase of solar metallicity populations (see M05). These models give spectra that embrace the variation in upturn strength modelled locally (MT00, Figure 5), but we consider a higher fuel consumption in order to account for the fact that BOSS galaxies are several Gyrs younger than local galaxies for which the upturn has been constrained. The fuel in the two coolest temperatures is a factor 3 larger than the strongest upturn galaxy in the local sample (NGC 4649). The fuel in the hottest temperature model is instead on the lower side of the local observations (see Figure \ref{fig:upturn_limits}). The upturn phase is then placed in each model with identical fuel and temperature independently of age without assuming any further time variation between 3 and 5 Gyr, which would be arbitrary.

Figure \ref{fig:upturns} shows the mid-UV region of a 5 Gyr population, for each of the upturn models described above. For comparison, we also include two models without a UV upturn contribution, named old - no upturn and old - no upturn - 2Z for the metallicities $Z_{\odot}$ and $2Z_{\odot}$, respectively. The absorption features seen in the models become more pronounced at higher temperature models, with the same amount of fuel, with the flux at the bluer end of the mid-UV decreasing. The continuum shows a much flatter shape for lower temperature models. The models with lower fuel show more similar shapes and features to models without an upturn contribution, as expected. Note that while the models extend to shorter wavelengths, Figure \ref{fig:upturns} uses the exact range of wavelengths covered by our data, as we have no data to constrain the model parameters out of this range.

The model grid spans ages of 3 - 14 Gyrs in 1 Gyr intervals. A finer grid at 0.1 Gyr intervals is obtained via interpolation by fitting a simple second or third order polynomial. 

It is important to note that the range of our explored models extends only mildly outside what is observed locally. In order to show this, we use the results of \citet{Maraston:2000} which modelled the upturns featured in the three massive local early-type galaxies NGC4472, NGC3379, NGC4649, and the bulge of M31. Figure \ref{fig:upturn_limits} is a remake of Figure 5 of MT00 and shows the upturn model  SEDs (with the same colour coding as in Figure \ref{fig:upturns}) normalised to 4699\AA. The y-intercepts of the weakest (NGC3379) and strongest (NGC4649) upturns modelled in MT00 are shown by the dash-dot and dotted lines respectively. Upturn models with lower amounts of fuel agree with these limits, with those featuring higher amounts falling above the range covered by the local galaxies. Models without an upturn component fall below the local range as expected.

\subsection{Mid-UV Absorption Features and Models}
\label{sec:line}

In this paper we focus on the mid-UV indices (2200 - 3200 \AA\ rest frame) due to the wavelength coverage and the redshift distribution of the observed spectra (3600 - 10,400 \AA\ observed frame, with an average $z \sim 0.57$).

Figure \ref{fig:theo_spec} shows the mid-UV region of a 1 Gyr population model with solar metallicity at the original resolution (before smoothing) with a downgraded spectrum overplotted in red.

Table \ref{tab:ind} recalls the index definitions of the 8 mid-UV features used in this work and provides relevant comments about the contributing atomic species according to the literature (see M09 for referencing).

The trends of these indices in young populations can be seen in Figure \ref{fig:trends}, adapted from M09. The figure shows the time evolution of synthetic line indices of simple stellar populations (SSPs) for different metallicities ($2 Z_{\odot}, Z_{\odot}, \frac{1}{2}Z_{\odot}$ and $\frac{1}{10}Z_{\odot}$). The dependence of the indices on age is straightforward, with all indices evolving strongly with age due to the fast evolution of the temperature of the turnoff stars at these low ages. The effect of metallicity becomes evident at ages larger than $\sim$ 100 Myr.

\setcounter{table}{1}
\begin{table*}
\begin{minipage}{126mm}
\caption{A list of indices used in this analysis, with bandpass definitions as denoted in \citet{Fanelli:1992}. Indices termed as "BL" are blends of several elements.}
\begin{tabular}{l || c | c | c | l}
Name & Blue bandpass & Central bandpass & Red bandpass & Comments \\ \hline
\hline
Fe II (2402\AA) & 2285.0 \ \ 2325.0 & 2382.0 \ \ 2422.0 & 2432.0 \ \ 2458.0 & \ \\
BL$_{2538}$& 2432.0 \ \ 2458.0 & 2520.0 \ \ 2556.0 & 2562.0 \ \ 2588.0 & Uncertain, Fe I? \\
Fe II (2609\AA) & 2562.0 \ \ 2588.0 & 2596.0 \ \ 2622.0 & 2647.0 \ \ 2673.0 & \ \\
Mg II & 2762.0 \ \ 2782.0 & 2784.0 \ \ 2814.0 & 2818.0 \ \ 2838.0 & \ \\
Mg I & 2818.0 \ \ 2838.0 & 2839.0 \ \ 2865.0 & 2906.0 \ \ 2936.0 & \ \\
Mg$_{wide}$ & 2470.0 \ \ 2670.0 & 2670.0 \ \ 2870.0 & 2930.0 \ \ 3130.0 & \ \\
Fe I & 2906.0 \ \ 2936.0 & 2965.0 \ \ 3025.0 & 3031.0 \ \ 3051.0 & \ \\
BL$_{3096}$ & 3031.0 \ \ 3051.0 & 3086.0 \ \ 3106.0 & 3115.0 \ \ 3155.0 & Al I 3092, Fe I 3091.6 \\
\label{tab:ind}
\end{tabular}
\end{minipage}
\end{table*}

A comparison of the UV indices measured on the individual IUE observed stars and the Kurucz stellar spectra as a function of temperature, with fixed gravity, was undertaken in M09. Several indices were found to be consistent between the real and synthetic stars, while others are sightly discrepant (e.g. BL1302, Si IV, BL1425, Mg II, Mg$_{wide}$, Fe I, and BL3096) . When this occurs the real stars often, but not always, have stronger indices. Some of these effects may originate from abundance ratio effects, others from an incorrect stellar parameter assignment or by incomplete line-lists in the theoretical model, non-LTE effects, or stellar winds.  Diffusion may also effect some of these indices, particularly those relating to iron, as it can cause overabundances to appear in the surface abundances.

When analysing globular clusters (GCs) in M09 the theoretical and empirical models were found to agree reasonably well for most indices in the far-UV, the exceptions being BL1425 and Fe V 1453 for which the empirical models were above the GC data. 

\setcounter{figure}{2}
\begin{figure*}
\includegraphics[width=0.8\textwidth]{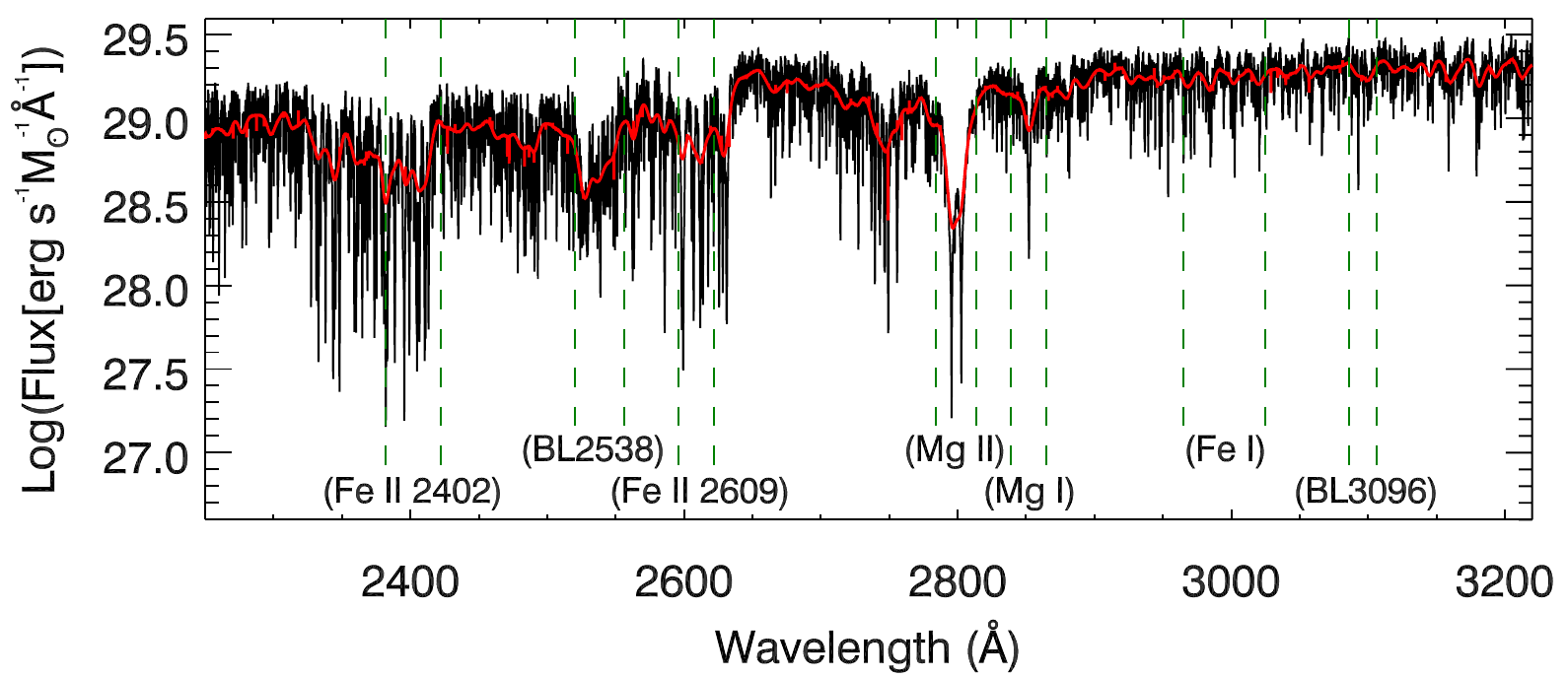}
\caption{The mid-UV region of a high resolution, theoretical M09 spectrum for a 1 Gyr, solar metallicity population. A smoothed (3\AA \  resolution) version of the spectrum, has been overplotted in red. The central bandpasses for the mid-UV indices are denoted by the dashed green lines. It should be noted that the Mg$_{wide}$ bandpasses are not shown to avoid crowding.}
\label{fig:theo_spec}
\end{figure*}

\setcounter{figure}{3}
\begin{figure*}
\includegraphics[width=0.85\textwidth]{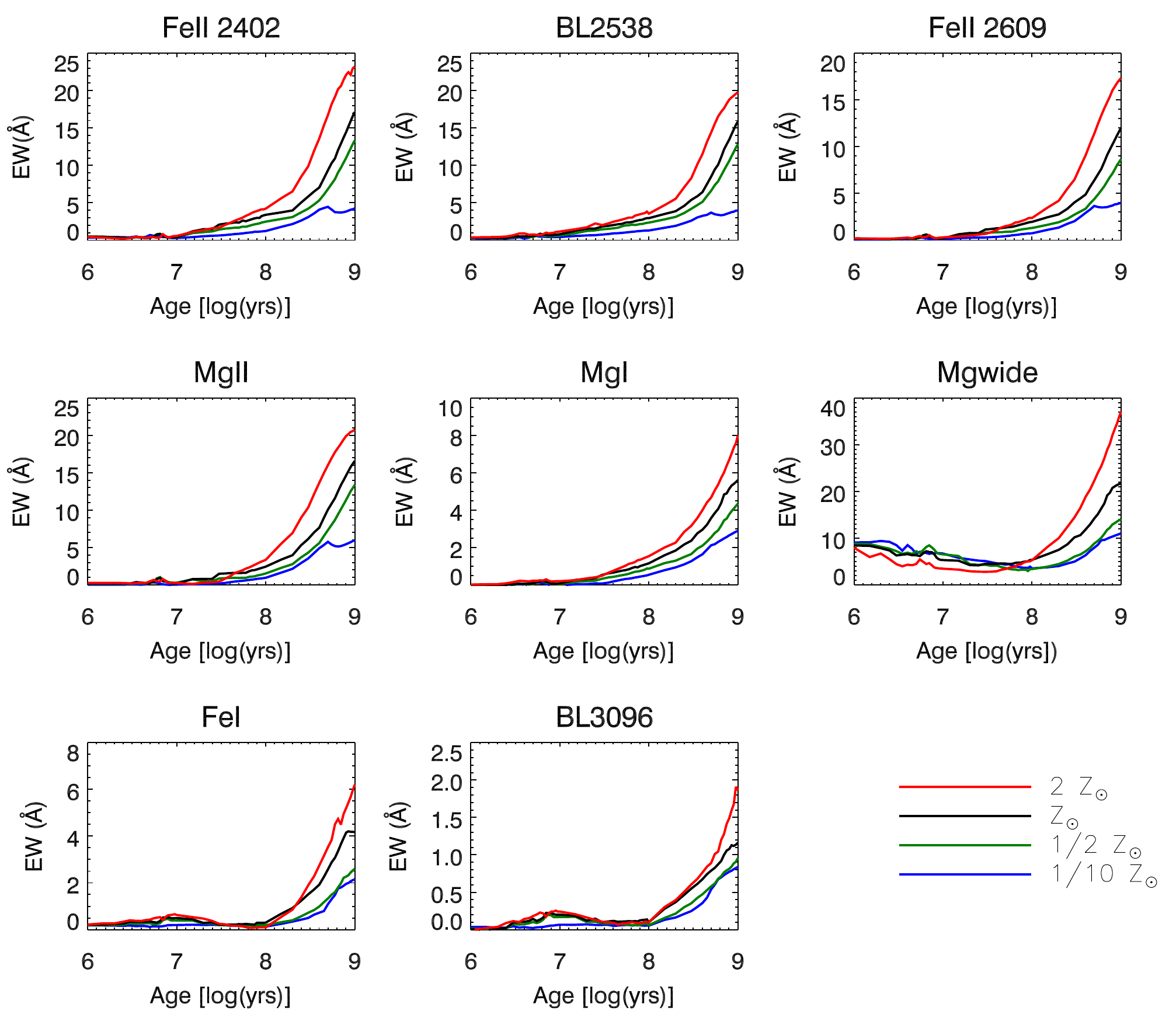}
\caption{Effect of age and metallicity on line indices of the theoretical SSP models. Adapted from \citet{Maraston:2009}, Figure 8.}
\label{fig:trends}
\end{figure*}

\setcounter{figure}{4}
\begin{figure*}
\subfloat[The effect of old age on the line indices of the theoretical $Z_{\odot}$ SSPs. The blue stars show the original M09 model, as shown by the solid black line in Figure \ref{fig:trends}, with the bhb extension including old UV-bright stars. The green triangles, cyan stars, cyan pluses, and light green squares show the T25, T40L, T40, and T10 upturn models respectively, with the red stars showing the old model without an upturn contribution.]{%
\includegraphics[width=0.72\textwidth]{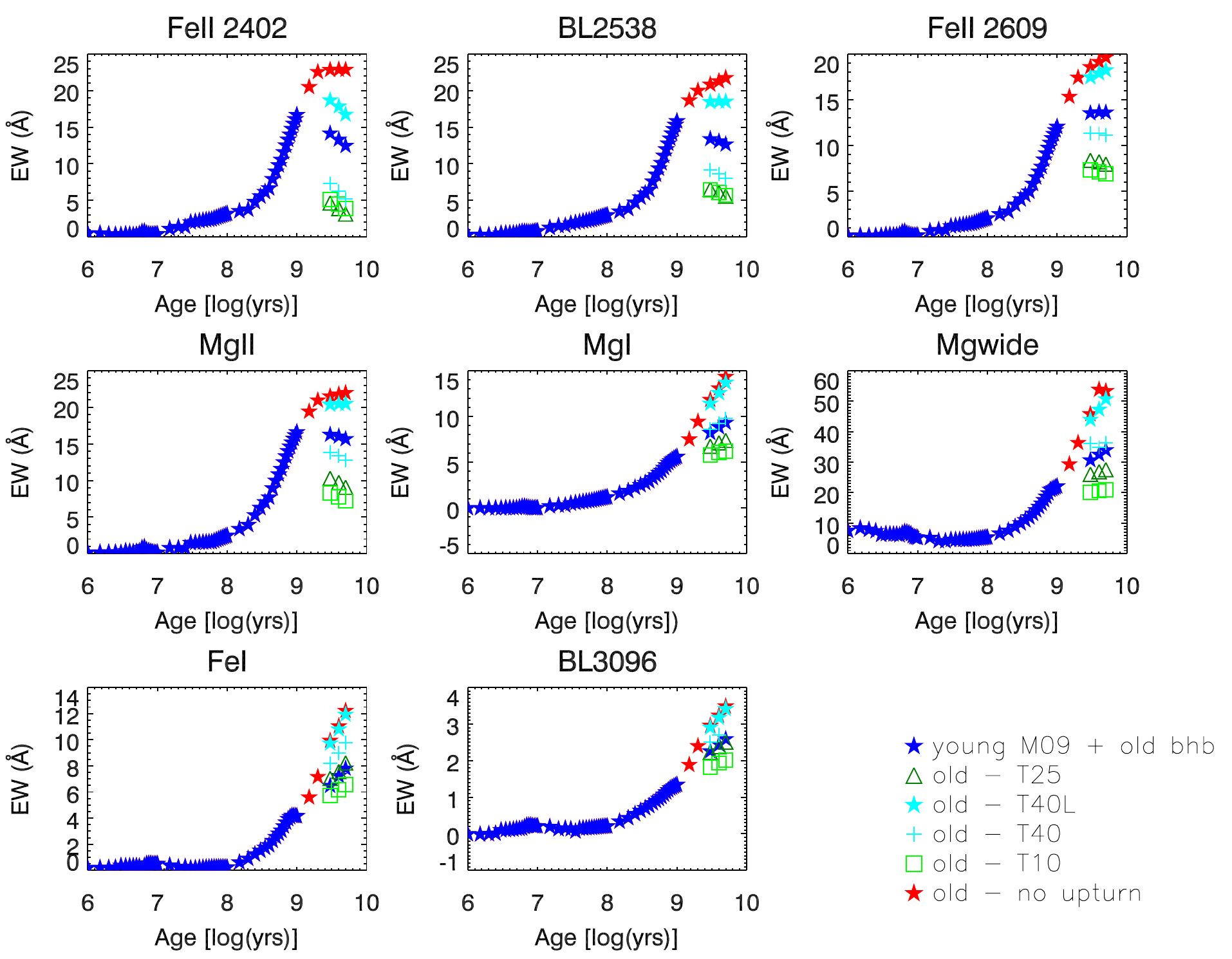}}\\
\subfloat[The effect of old age on the line indices of the theoretical $2Z_{\odot}$ SSPs. The blue stars show the original M09 model ($Z_{\odot}$), as shown by the solid black line in Figure \ref{fig:trends}. The love triangles and turquoise stars show the T25\_2Z and T25L\_2Z models with the purple squares and black pluses the T35\_2Z and T35L\_2Z models. The red asterisks show an old model of the same metallicity without an upturn contribution.]{%
\includegraphics[width=0.72\textwidth]{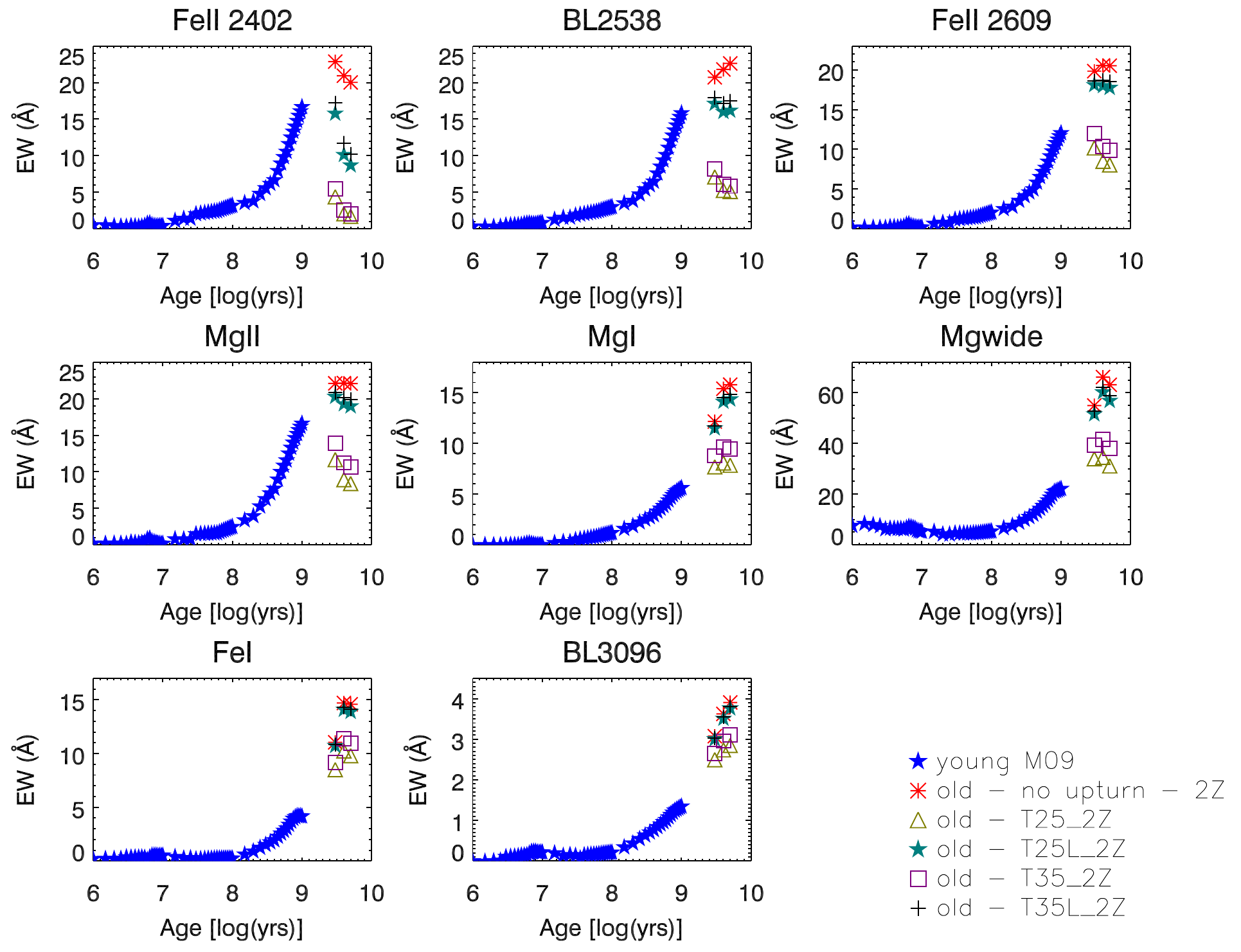}}
\caption{The effect of old age on the line indices of the theoretical SSP models. \textit{Top panel:} $Z+{\odot}$ models. \textit{Bottom panel:} $2Z_{\odot}$ models.}
\label{fig:ex_trend}
\end{figure*}

\setcounter{figure}{5}
\begin{figure*}
\begin{minipage}[b]{0.45\textwidth}
\includegraphics[width=\textwidth]{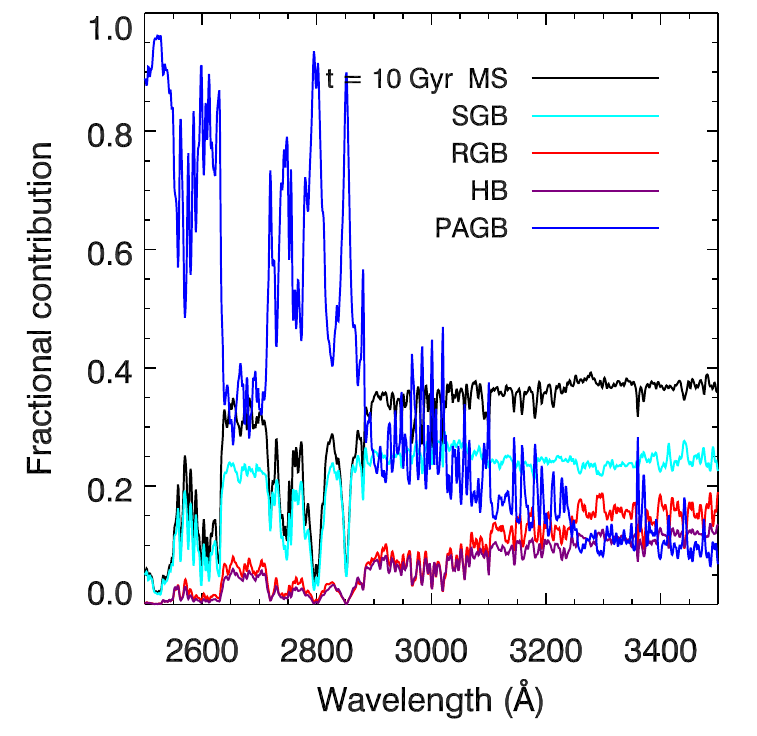}
\end{minipage}
\begin{minipage}[b]{0.45\textwidth}
\includegraphics[width=\textwidth]{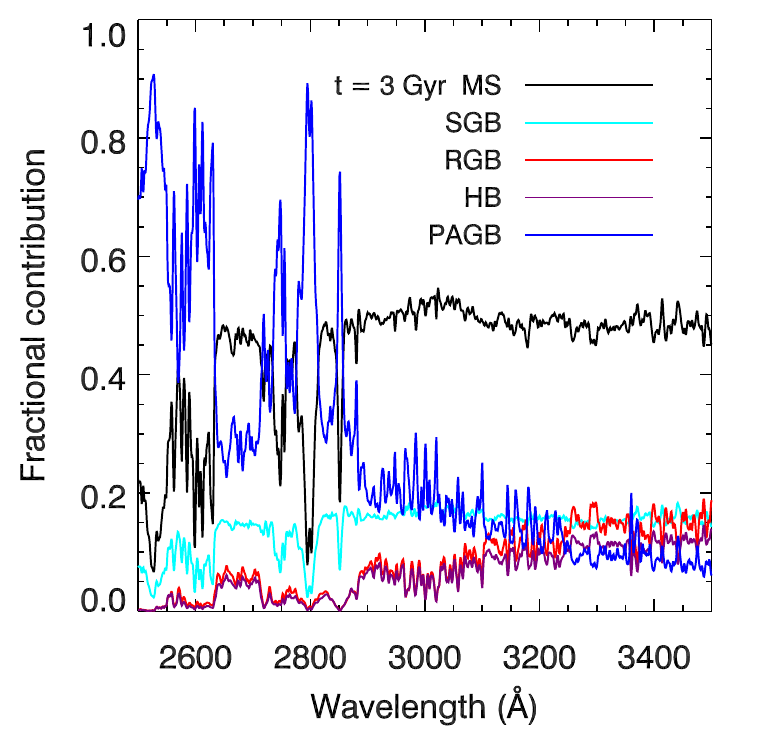}
\end{minipage}
\caption{The fractional contribution of evolutionary phases to the mid-UV spectral region of a 10 Gyr (\textit{LHS})and 3 Gyr (\textit{RHS}) population. The blue line shows the contribution from UV bright (PAGB) stars and the black from main sequence (MS) stars. The cyan and red lines shows the contributions from the sub-giant brach (SGB) and the red giant branch (RGB) with the contribution from a red  horizontal branch (HB) stars shown by the magenta line.}
\label{fig:cont_frac}
\end{figure*}

In the case of the mid-UV indices, Fe II 2402 and Mg$_{wide}$ were found to agree well, however the validation of these indices is less conclusive due to the ages of the GCs being limited to 130 Myr and the requirement of older ages to set constraints on the mid-UV region. Interstellar lines are known to effect the region around the magnesium lines in the mid-UV therefore they are the most likely reason for the discrepancy but one could also argue that element abundance ratio effects start affecting the indices in the mid-UV.

The effect of old UV-bright populations on the mid-UV indices can be seen for all of the upturn models in Figure \ref{fig:ex_trend}, with the $Z_{\odot}$ models shown in (a) and the $2Z_{\odot}$ models in (b). The EWs of Fe II 2402, BL2538, Fe II 2609, and Mg II decrease in value, or begin to plateau, with the addition of old UV-bright populations creating a degeneracy between young and old UV-bright populations. Mg I, Mg$_{wide}$, Fe I, and BL3096 all continue to increase in strength with increasing age making them non-degenerate. These indices are potentially powerful to investigate the two populations.

The turnover in EW strength seen in the first four indices is not present in the old model without a contribution from UV-bright stars, denoted in Figure \ref{fig:ex_trend} by the red stars. Hence these indices can be used to differentiate between an old population with and without the presence of UV-bright stars.

Figure \ref{fig:cont_frac} shows the fractional contributions of different evolutionary phases to the mid-UV region that covers the non-degenerate indices (Mg I, Fe I, and BL3096) for both a 10 Gyr and 3 Gyr population, \textit{LHS} and \textit{RHS} respectively. UV-bright stars (PAGB), the solid blue line, contribute a sizeable amount of light around the region of the Mg I line ($\sim$ 2700 - 2900\AA). Together with main sequence stars, shown by the solid black line, they are the major contributors. At younger ages, shown on the \textit{RHS}, the PAGB contribution is somewhat lower towards the redder end of the mid-UV, but is still relevant.

\section{SDSS-III / BOSS Galaxy Data}
\label{sec:data}

The observed galaxy spectra come from the Sloan Digital Sky Survey (SDSS) III's Baryon Oscillation Spectroscopic Survey (BOSS) (\citet{Dawson:2013}) which has mapped $\sim1.5$ million massive, luminous, galaxies out to $z \sim 0.7$, with an average of $z \sim 0.57$. We use Data Release 12 (DR12) which contains  spectra from all previous data releases of BOSS as well as all imaging and spectra from prior SDSS data releases.

In the BOSS sample, due to a combination of the wavelength coverage of the spectra (3600 - 10,400\AA) and the redshift distribution of the sample we have access to the mid-UV indices above $z = 0.6$. 

Figure \ref{fig:BOSS_spec} (a) shows a typical BOSS spectrum taken from our working sample, with $z$ = 0.61 and an average signal to noise ratio (SNR) of 0.65 over 2250 - 3200\AA.

\setcounter{figure}{6}
\begin{figure*}
\centering
\subfloat[Mid-UV indices for a typical BOSS spectrum]{%
\includegraphics[width=0.65\textwidth]{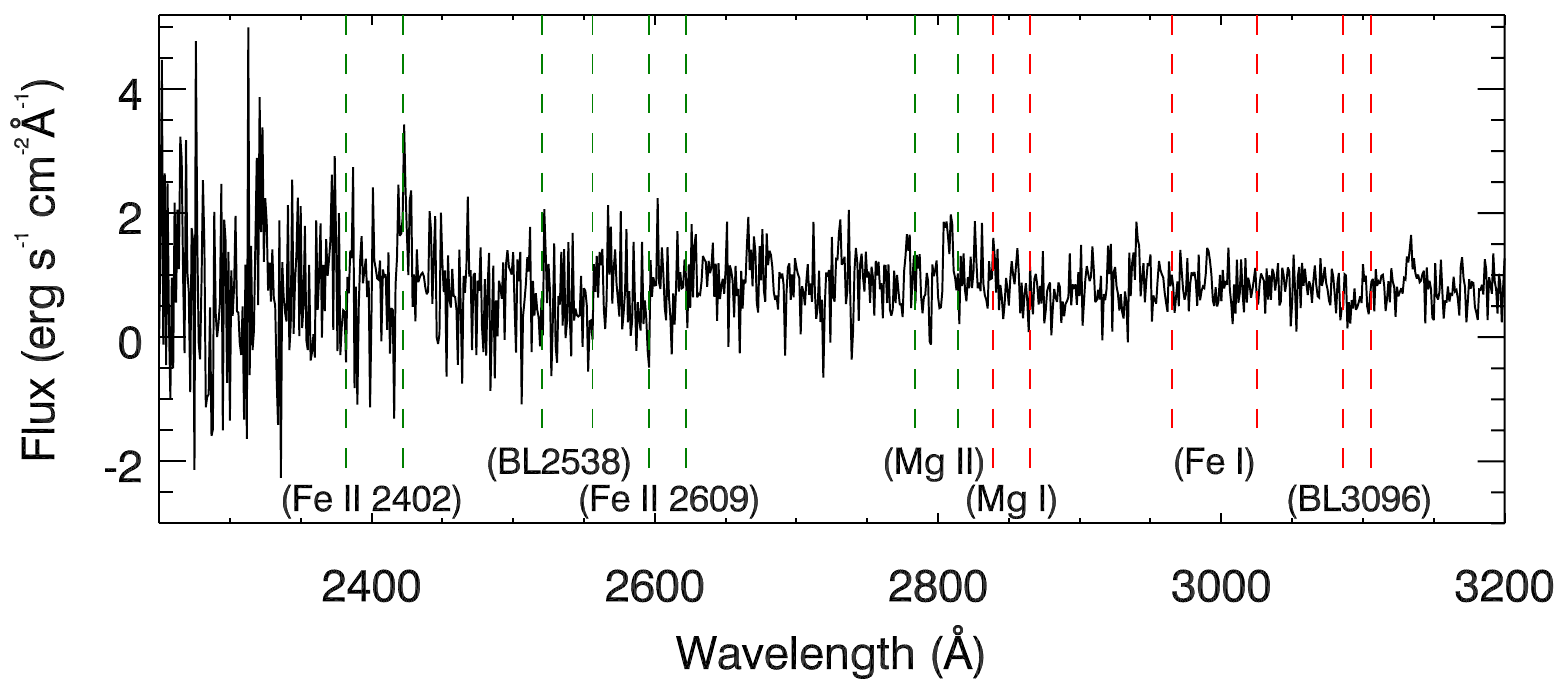}}\\
\subfloat[The mid-UV indices for a set of higher SNR stacks]{%
\includegraphics[width=0.65\textwidth]{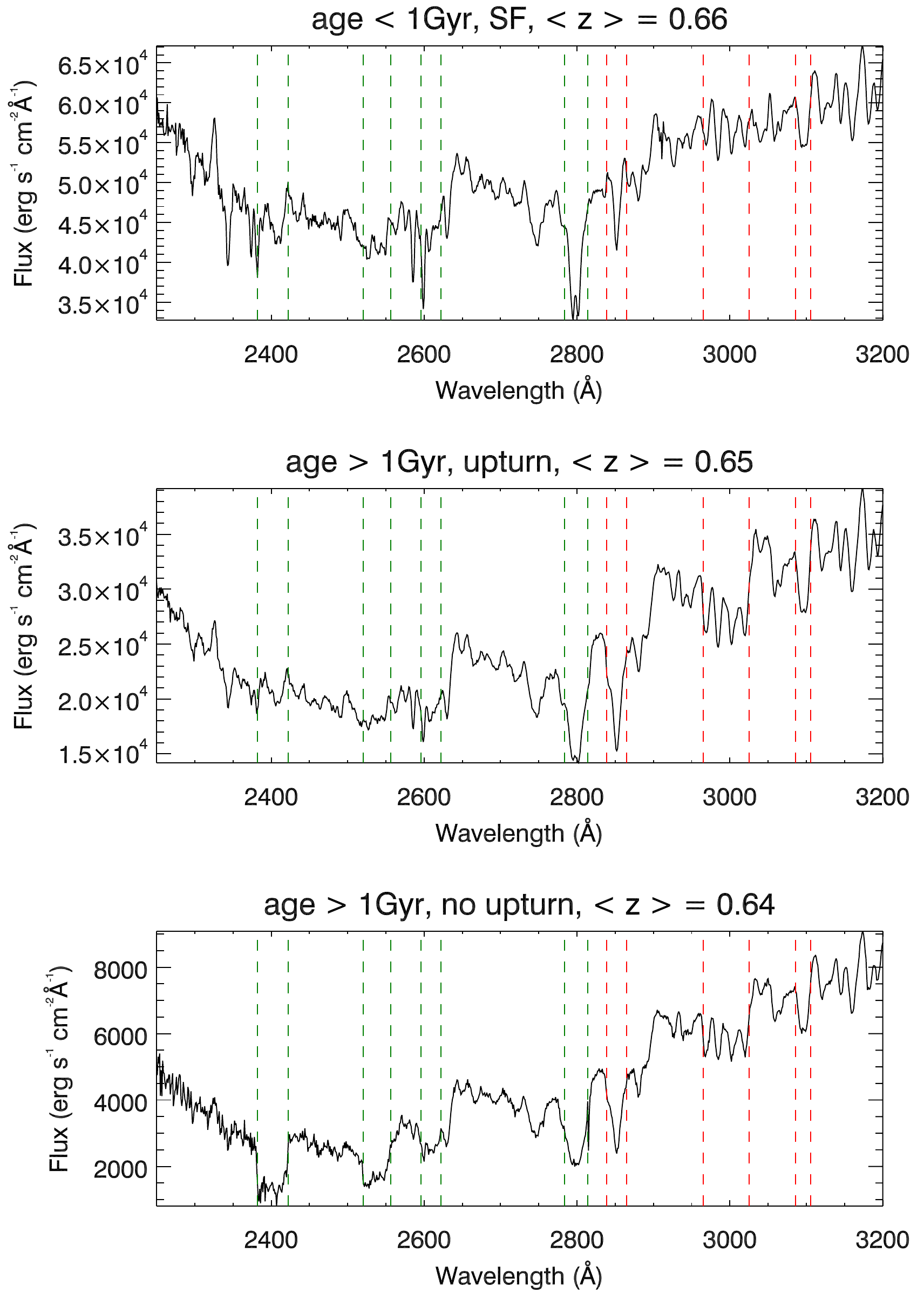}}
\caption{(a) The mid-UV region of a typical BOSS spectrum, z = 0.61 and SNR of 0.64, over 2250 - 3200\AA. The central bandpasses for the degenerate mid-UV indices are denoted by the dashed green lines, with the non-degenerate in red. It should be noted that the Mg$_{wide}$ bandpasses are not shown to avoid crowding. (b) The mid-UV region of a set of higher SNR stacks. Central bandpasses as in (a)}
\label{fig:BOSS_spec}
\end{figure*}

\

These BOSS spectra have not been "cleaned" of emission lines below 3600\AA. Although only a small fraction ($\sim 5\%$) have emission (\citet{Thomas:2013boss}), care must still be taken when dealing with indices which may contain contamination from such lines as the stellar population models used in this work do not include emission processes. For example the Mg$_{wide}$ index is known to have potential contamination due to several emission lines which fall within it's central bandpass (Christy Tremonti, \textit{private communication}). It is for this reason that we exclude the Mg$_{wide}$ index from much of our analysis.

In the following subsections we discuss some of the operations applied to the observed spectra as well as additional sources of data such as the BOSS galaxy products from which galaxy properties are extracted.

\subsection{BOSS Galaxy Properties}
\label{subsec:product}

We shall further use results from the so-called Portsmouth galaxy product\footnote{\textit{www.sdss3.org/dr10/spectra/galaxy\_portsmouth.php}} in which stellar masses, ages, star formation rates, and metallicities are derived from broadband SED fitting of the observed $ugriz$ magnitudes of BOSS galaxies with the spectroscopic redshift determined by the BOSS pipeline (\citet{Bolton:2012}). The best-fit is obtained using various stellar population models;

\begin{enumerate}
\item a passively evolving galaxy with no ongoing star formation with a two component metallicity of the same age, as in \citet{Maraston:2009colour}
\item one allowing new stars to form, for various metallicities and timescales, with an ensemble of star formation modes including; exponentially-declining, constant with truncation, and constant star formation, as in \citet{Maraston:2006}. \end{enumerate}

These two sets of template spectra provide two sets of properties for each BOSS galaxy. 

Galaxy parameters including emission line fluxes, stellar and gas kinematics, and velocity dispersions have been calculated for each galaxy and published as BOSS galaxy products in \citet{Thomas:2013boss}, with the calculated stellar masses and ages in \citet{Maraston:2013boss}.

\

\citet{Masters:2011} show that using a simple colour cut of ($g - i \ge 2.35$) allows one to be able to select a sub sample of BOSS galaxies with $\ge$ 80\% early type morphology. The remaining 20\% of galaxies above this cut have a late type morphology. This classification scheme allows us to study the difference between the UV contributions found in passive galaxies and those more likely to have ongoing star formation.

\subsection{Stacking Spectra}
\label{subsec:stacks}

Stacks were created, binning in $g-i$ colour, redshift, velocity dispersion, and UV age to provide higher SNR spectra to help constrain the quality and understand which features can be measured.  The stacks were produced using the stacking code of \citet{Thomas:2013boss}, taking the values of redshift and velocity dispersion produced in their analysis of BOSS galaxies.

The galaxies were split in colour by a cut at $g-i < 2.35$ as well as a divide in the derived UV age (see Section \ref{sec:analysis} for a description of how we derive UV ages) to split young populations as well as old populations with and without a contribution from UV-bright stars. We also created three stacks, known as master stacks throughout the paper, stacking only in UV age and ignoring all other stacking selections, to create a master stack for each of the different age populations (young, old with upturn, and old without upturn).

We correct for dust as follows. Spectra are corrected for Milky Way foreground dust reddening by extracting $E(B - V)$ values for each object (\citet{Schlegel:1998}) and calculating the attenuation by utilising the \citet{O'Donnell:1994} extinction curve. Non Milky Way reddening is corrected for by adopting the dust law of \citet{Calzetti:2000}.

The stacks were created by summing the flux of BOSS galaxies in selected parameter bins and resampling them into a linear wavelength grid covering 2250 - 6200\AA, rest frame, in steps of 1\AA. Galaxies were required to have a minimum SNR of 1.5pix$^{-1}$ and a maximum error in velocity dispersion of 30\% to minimise contamination between velocity dispersion bins.

A comparison of the detail seen in the different SNRs can be seen in Figure \ref{fig:BOSS_spec}, which shows a typical BOSS spectrum ( $< z > = 0.61$, SNR = 0.64) in panel (a), with the three master stacks for each age population shown in panel (b). 

Firstly, it should be noted that the absorption features become much more evident in the stacks. The Mg I, Fe I, and BL3096 features are stronger in the upturn selected stack than in the young stack. These three indices are also more prominent in the old stack without an upturn however the other four indices are also much broader and more pronounced than in both the young and upturn stacks.

It should be noted that an increase in flux below 2400\AA \  is seen in all three of the master stacks. This spectral shape occurs at approximately the same observed wavelength in all stacks making this feature a systematic of the data rather than a true upturn. We also note that there is significant broadening of the bluer indices (shortward 2600\AA) in the stack created from galaxies selecting old ages without an upturn component. This may be due to the blending of other nearby absorption features that are not present in hot stars. We shall investigate the source of broadening in future work, however this seems like an interesting discriminator of an old population devoid of hot components.

\section{Qualitative Matching of Models and Data, and Error Analysis}
\label{sec:cover}

We begin by comparing the EWs of all 8 mid-UV indices as calculated for each individual spectrum and stack in our working sample with the range encompassed by the models.

The error on the EW of each index was estimated by perturbing the flux at each wavelength for each spectrum by a random amount. This perturbation was calculated by taking a random number from a normal distribution between 0 and 1 for each wavelength and multiplying it by the associated error range given by BOSS for that wavelength point of the spectrum. 

This process was performed 1000 times for each spectrum, producing a spectrum with slight perturbations from the original each time. For each perturbed spectrum produced, the EW was recalculated and the distribution of the values obtained from a sample individual spectrum for each index can be seen by the red distributions in Figure \ref{fig:mc_error}. The peak of the distribution is centred around the value calculated from the original sample spectrum. An estimate of the error on each index was yielded by fitting a Gaussian to the distributions and calculating the standard deviation. 

\setcounter{figure}{7}
\begin{figure*}
\includegraphics[width=0.86\textwidth]{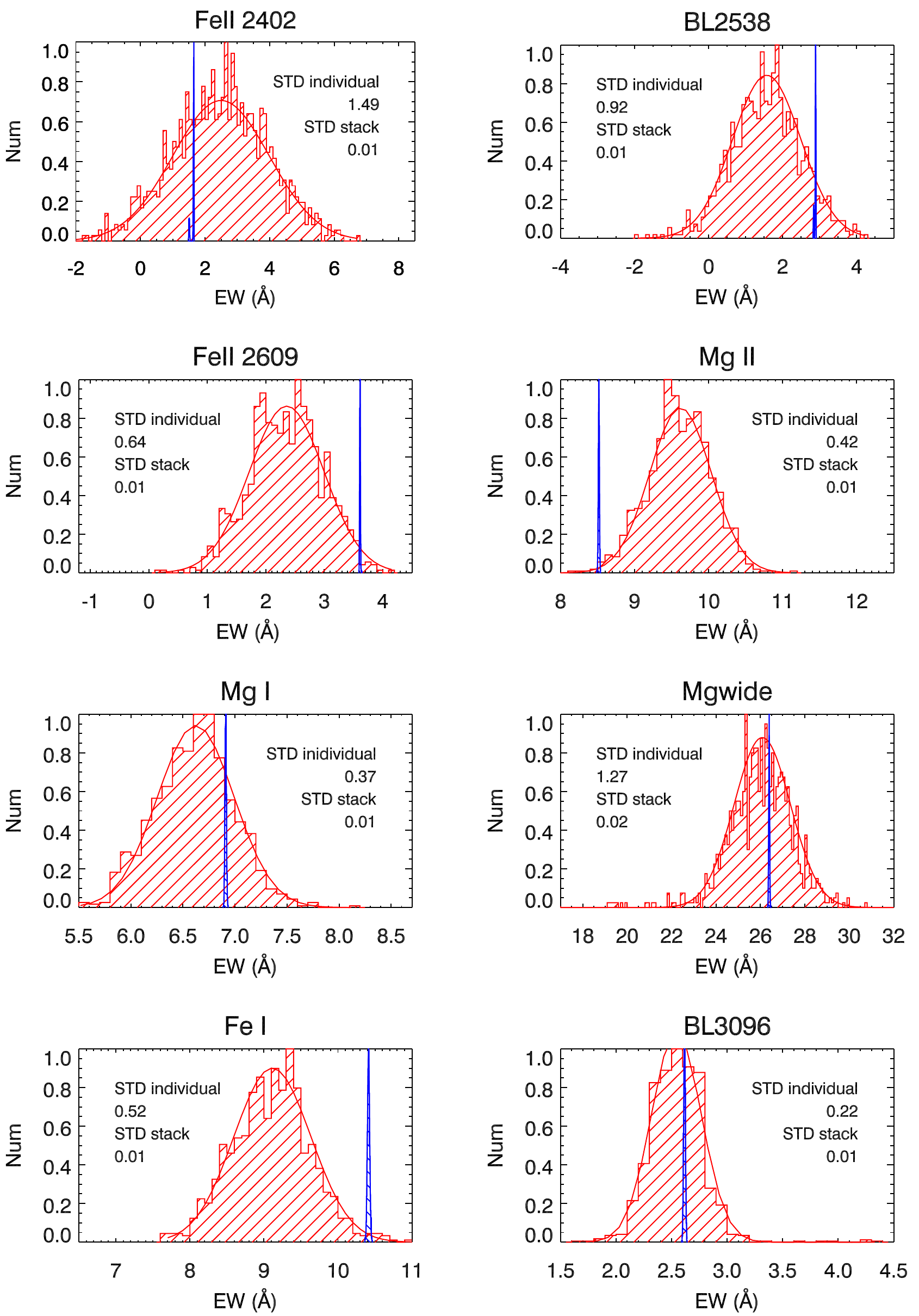}
\caption{Distribution of EWs calculated from perturbing a sample spectrum 1000 times, selecting random errors within the given range for each pixel. The solid red line shows the gaussian fit for each distribution  from which the standard deviation is calculated. The blue shows a higher SNR stack. Each distribution has been normalised to 1. The standard deviations calculated for each index for both the individual and stacked spectra can be seen in the respective panels.}
\label{fig:mc_error}
\end{figure*}

The same analysis was performed on each stack yielding significantly smaller errors for each index, an example of which can be seen by the blue distribution in Figure \ref{fig:mc_error}. This holds true for the majority of the errors calculated for all the individual BOSS spectra and stacks. The standard deviations calculated for each index, for both the
individual sample spectrum and stack, can be seen in the respective index panels.

These errors can be compared to the difference between models at older ages to determine if we can resolve the effect seen in the models in our data.

Figure \ref{fig:error_comp} shows the difference in EW between the models with an old UV-bright component and the one without (see Figure \ref{fig:ex_trend}) for each model age in comparison to the estimated errors on the calculated EW for both an individual spectrum, blue solid line, and a stack, green solid line, with the dashed lines showing twice the standard deviation. The estimated errors are those from Figure \ref{fig:mc_error}.

\setcounter{figure}{8}
\begin{figure*}
\subfloat[The difference between $Z_{\odot}$ theoretical models with and without an upturn contribution. See the legend for model coding.]{%
\includegraphics[width=0.9\textwidth]{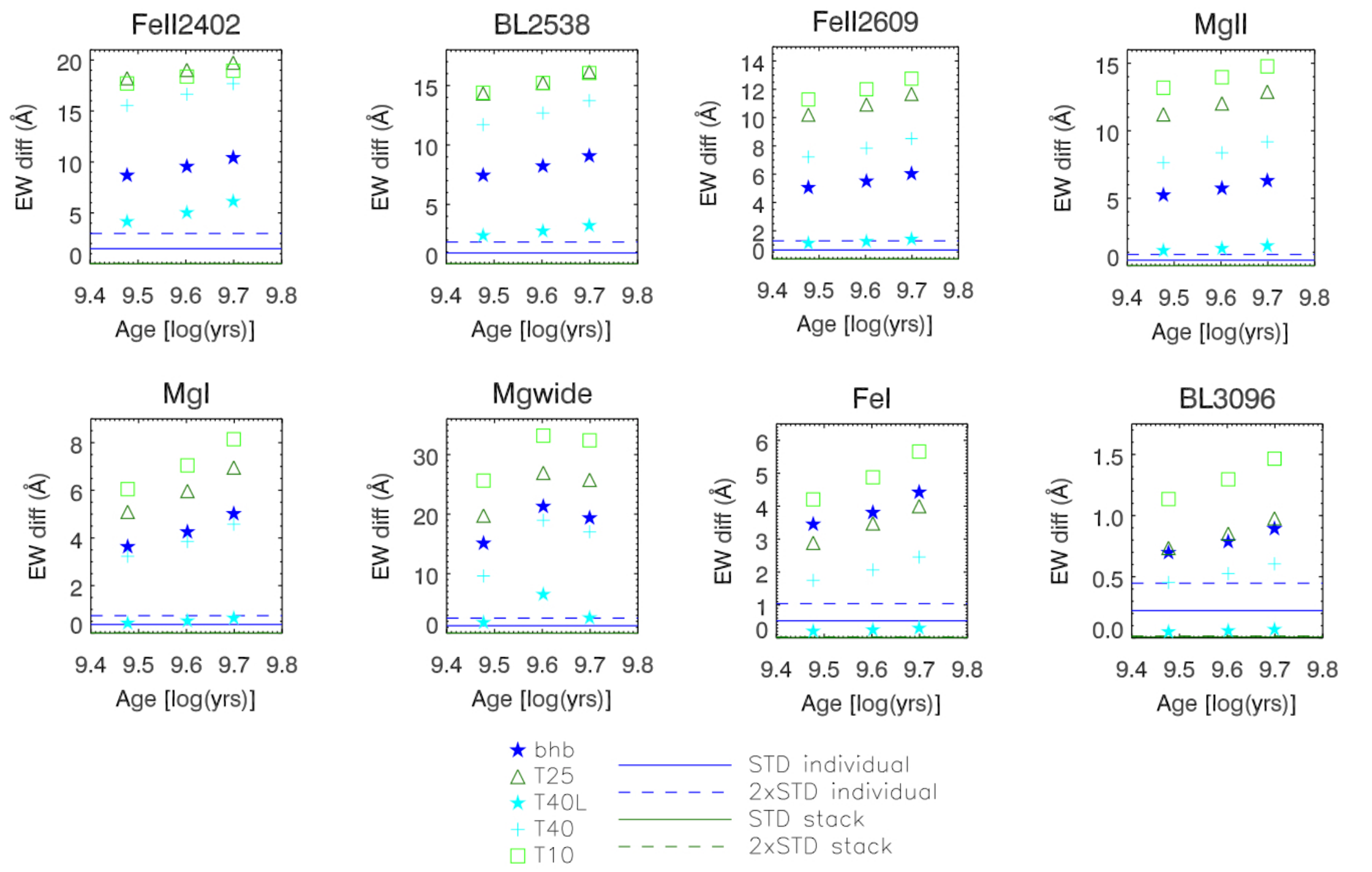}}\\
\subfloat[As in (a) but with $2Z_{\odot}$ theoretical models. See the legend for model coding.]{%
\includegraphics[width=0.9\textwidth]{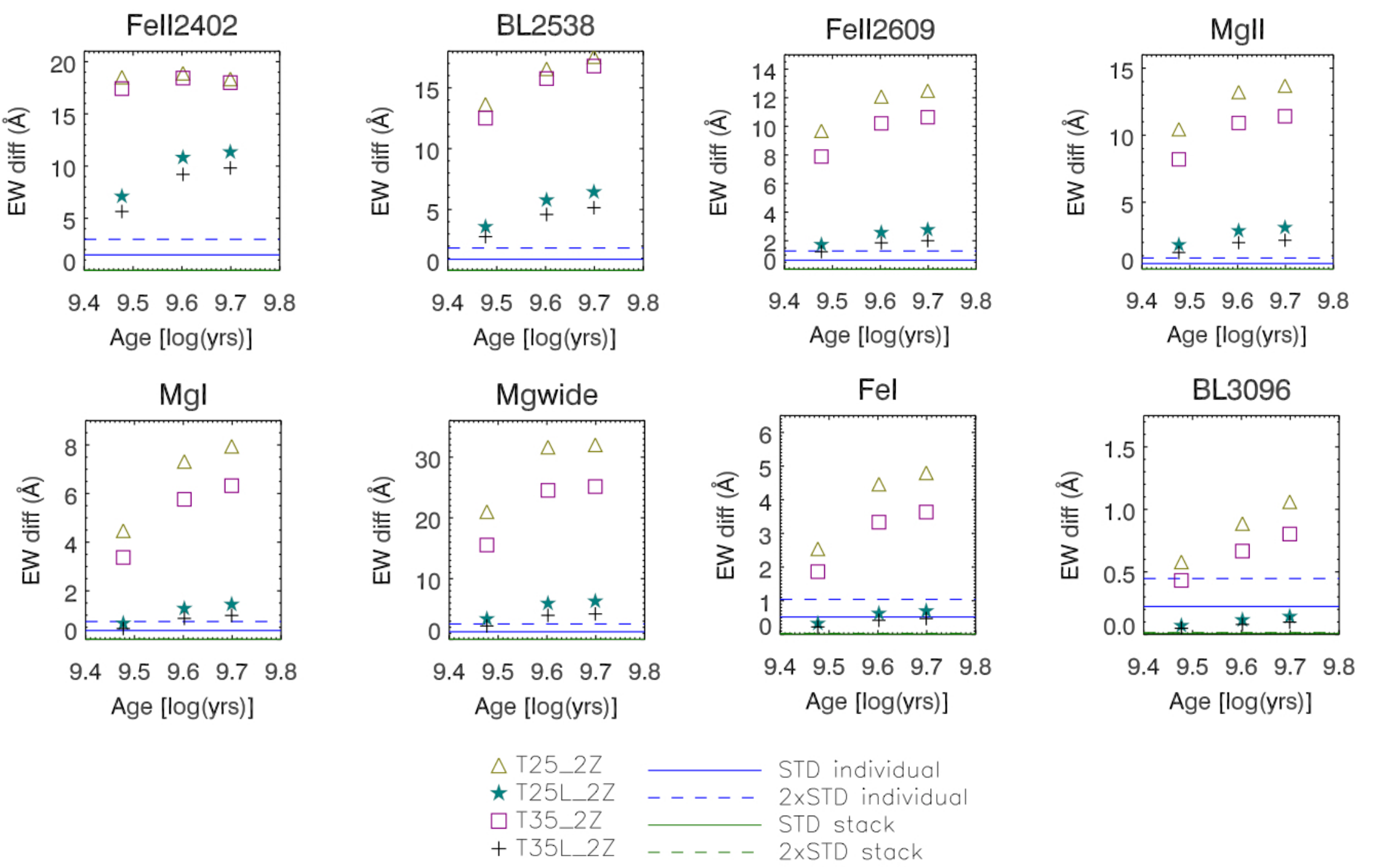}}
\caption{The difference in EW for models containing a contribution from old UV-bright stars and an old model without UV-bright stars. The solid blue and green lines denote the standard deviation of the EW for a sample individual spectrum and a stack (see Section \ref{sec:cover}) respectively. The dashed blue and green lines shows twice the standard deviation.}
\label{fig:error_comp}
\end{figure*}

For most indices the difference in EW for ages above 3 Gyr (log(age [yrs]) $\sim$ 9.48) is greater than twice the error on the EWs in our data, with the exception of the models featuring lower amounts of fuel (T40L in (a) and T25L\_2Z and T35L\_2Z in (b)) which fall around or below one standard deviation. As discussed above, the error on the EW of the stacks is significantly lower than that of the individual with one standard deviation falling far below the EW difference for all indices at ages above 3 Gyrs. Therefore we should be able to see this effect in our data.

\

Figure \ref{fig:coverage} shows the range of EW values calculated from our working sample with the errors associated with each index shown by the error bars, on either side of the cross symbols, in the upper righthand corner of each panel. Individual spectra are shown in the red histogram and stacks in black, both distributions have been normalised to one. The area in which our model values lie (both empirical and theoretical M09 models and all theoretical extensions)\footnote{\label{mgnote}N.B There is only a theoretical model for the Mg$_{wide}$ index.} is shown by the grey shaded area.

In general, the majority of the indices show good agreement between the data and the models. The individual spectra span a much larger range than the models, extending beyond the axes shown. A fraction of this spread can be accounted for by the errors on the EW calculation: however, there is still significant extension towards both stronger and weaker EW values. The EWs of the stacks mainly fall within the area covered by the models due to the increase in SNR. 

The extension of some EWs measured on stacks beyond the values found in the model is too large to be accounted for by the error on the EW alone. Stacks with EWs that fall above the model range are, in general, created by stacking galaxies which select old ages from the no upturn model which have the highest EW values. Those that fall below are created from galaxies selecting ages from the original, young M09 model with inherently smaller EWs.

In Figure \ref{fig:snr_z} we show how the redshift and signal to noise ratio over 2250 - 3200\AA \ (SNR$_{UV}$) affects whether the measured EWs agree with the values found in the models. To do this we calculate whether the measured EWs fall within the minimum and maximum values of the model range shown in Figure \ref{fig:coverage}.

The colours in Figure \ref{fig:snr_z} show the percentage of individual spectra that fall within our model range. The redshifts are those obtained via the BOSS pipeline and the SNR is calculated as the average SNR between 2250 - 3200\AA.

For the individual spectra, as redshift increases more galaxies with lower SNR agree with our model range due to the wavelengths under analysis being observed more in the optical range where the BOSS spectrograph is optimised. The Mg$_{wide}$ index is shown to have less agreement with the model range than the other indices, which could be due to contamination from emission lines. The discrepancy between the values found in the data and the range covered by our models strengthens our decision to exclude this index from further analysis.

The stacks follow the same trend as the individual spectra for the majority of indices with the most notable exception being Fe I. Several stacks fall outside of the EW range for each index the most notable being Fe II 2402, BL2538, Fe I, and BL3096. The vast majority of stacks which fall outside of the model range are those created stacking galaxy spectra selecting old ages from the no upturn model, with stacks from young ages and upturn models falling within the model range. In the case of Fe I, several of the stacks created from galaxies selecting young ages also fall outside of the model range with EWs lower than those found in the model range, as seen in Figure \ref{fig:coverage}.

\setcounter{figure}{9}
\begin{figure*}
\includegraphics[width=0.85\textwidth]{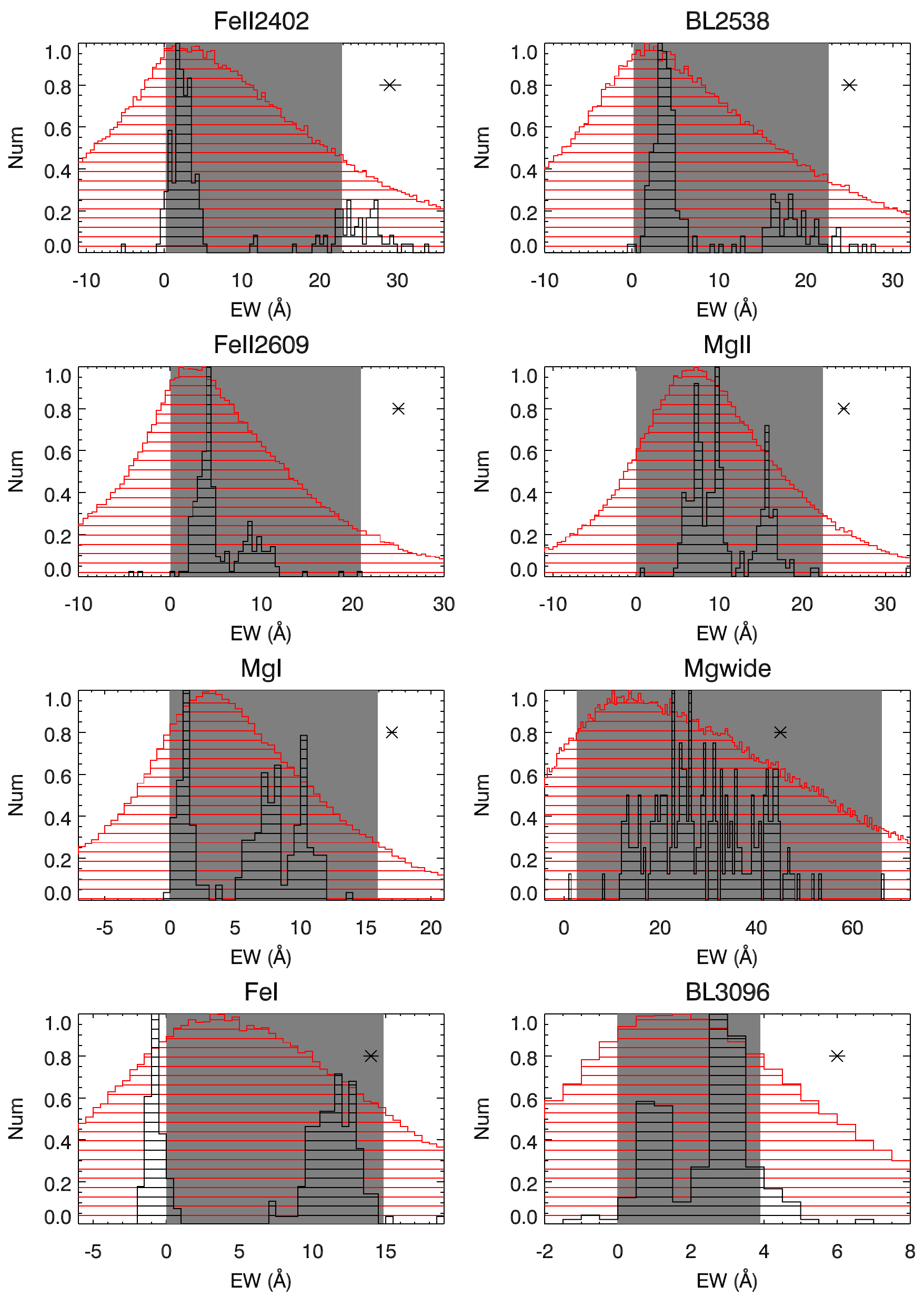}
\caption{Comparison of the distribution of EWs found in the individual spectra, red, and the stacks, black. The area in which our models lie  for each index is shown by the grey shaded area$^{\ref{mgnote}}$. The error on the EW for individual spectra, as calculated from Figure \ref{fig:mc_error}, is shown in the top right hand corner of each distribution by the lines extending horizontally from the cross symbols. Each distribution has been normalised to 1.}
\label{fig:coverage}
\end{figure*}

\setcounter{figure}{10}
\begin{figure*}
\includegraphics[width=0.7\textwidth]{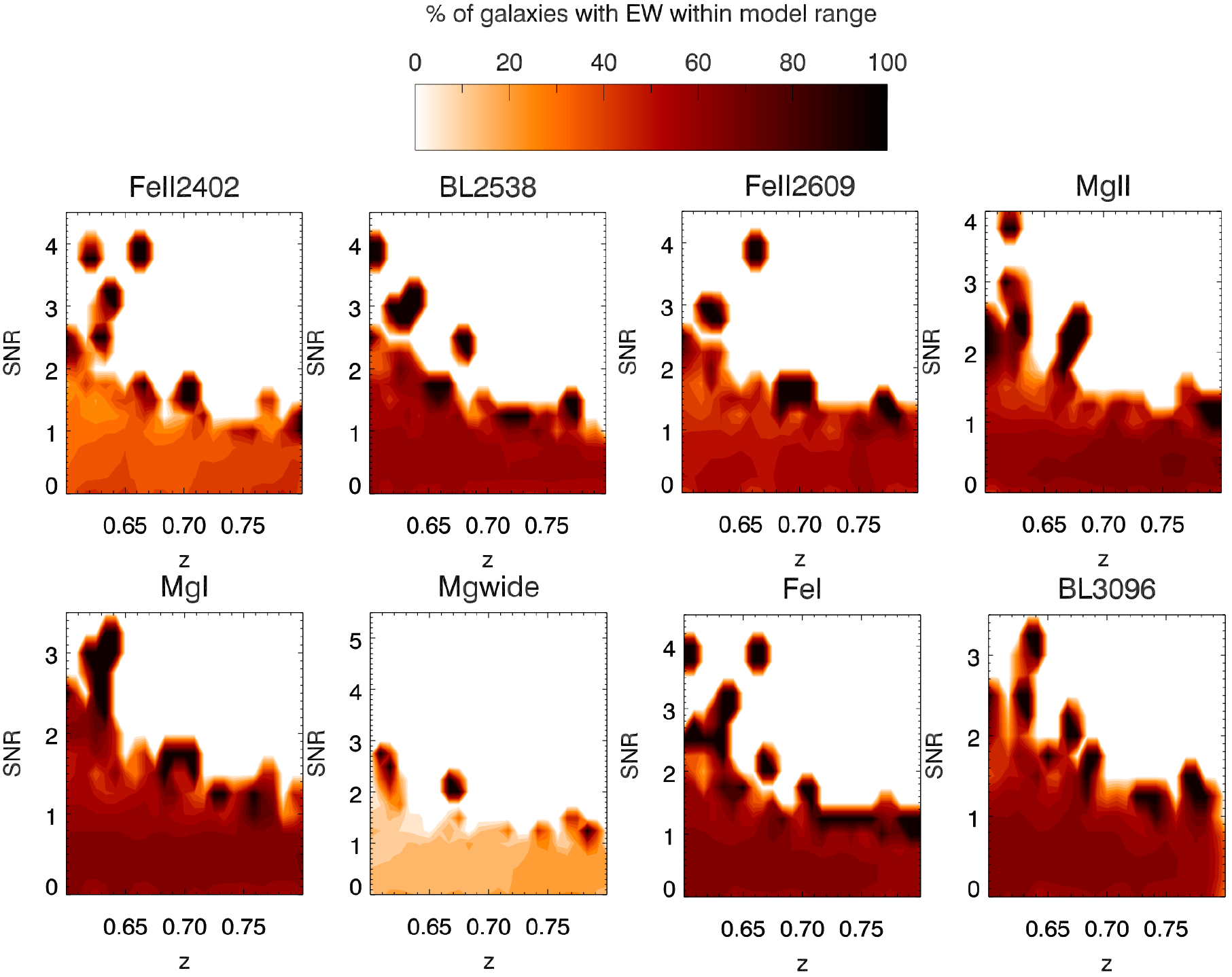}
\caption{The effect of redshift and SNR$_{UV}$ on whether the measured EW falls within the range of our models. The contours, showing individual spectra, give the percentage of galaxies that agree with the range of EWs allowed.}
\label{fig:snr_z}
\end{figure*}

The decrease in SNR seen towards higher redshifts in the stacks  is due to two factors. Firstly the SNR decreases naturally as high redshift galaxies become fainter due to cosmological dimming.  Secondly the number of spectra available to stack decreases as the redshift increases due to the redshift distribution of the survey, which peaks at $z \sim 0.57$ (\citet{Dawson:2013}). This is reflected in the SNR of the stacks. Those at lower redshifts, comprised of more individual spectra, have higher SNR ratios than those at higher redshifts, comprised of fewer individual spectra. It should also be noted that SNR of the individual galaxies is higher at lower redshifts, which also adds to the trend seen in the stacks.

\setcounter{figure}{11}
\begin{figure*}
\subfloat[Theoretical  $Z_{\odot}$ model grids. See legend for model coding.]{%
\includegraphics[width=0.9\textwidth]{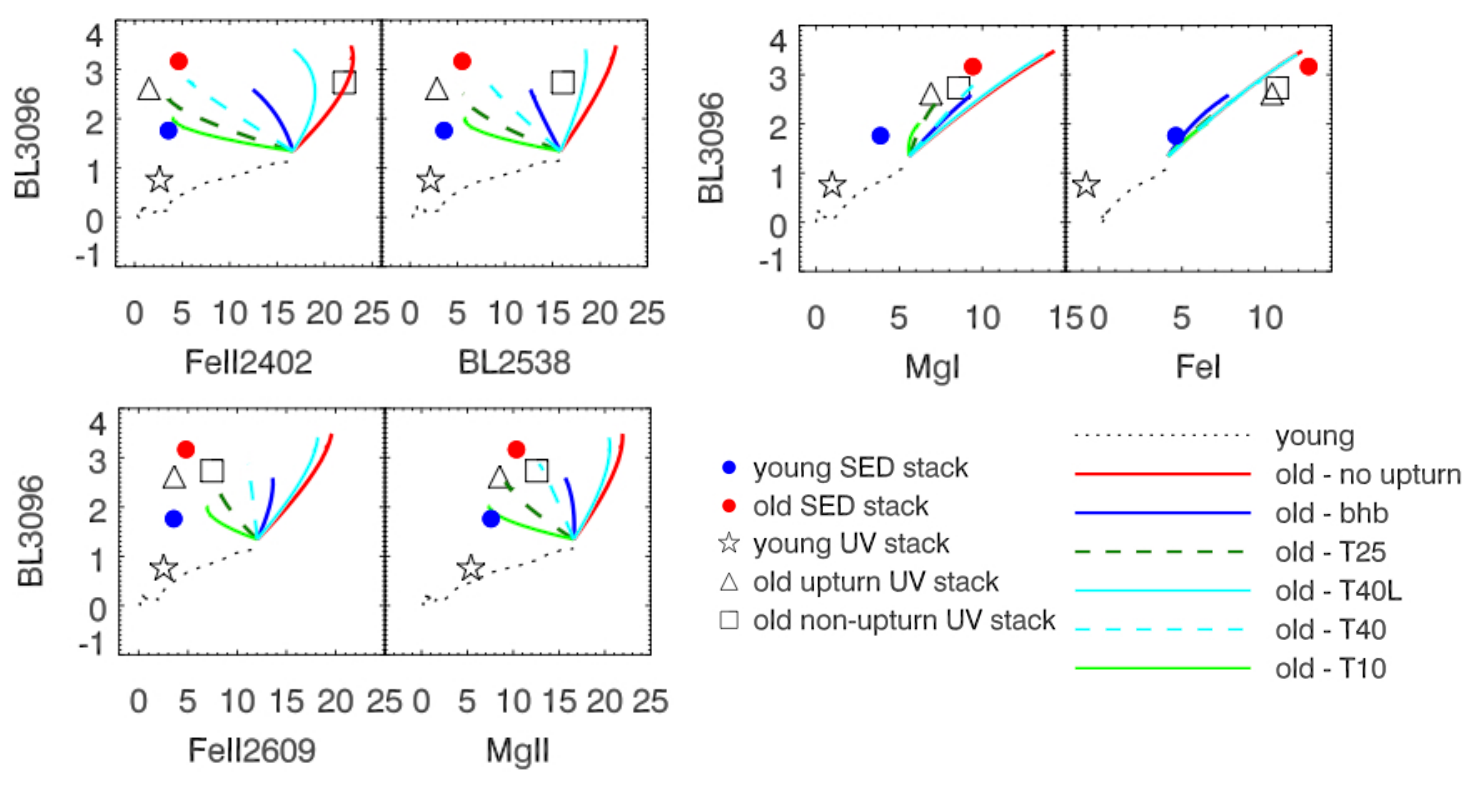}}\\
\subfloat[Theoretical $2Z_{\odot}$ model grids. See legend for model coding.]{%
\includegraphics[width=0.9\textwidth]{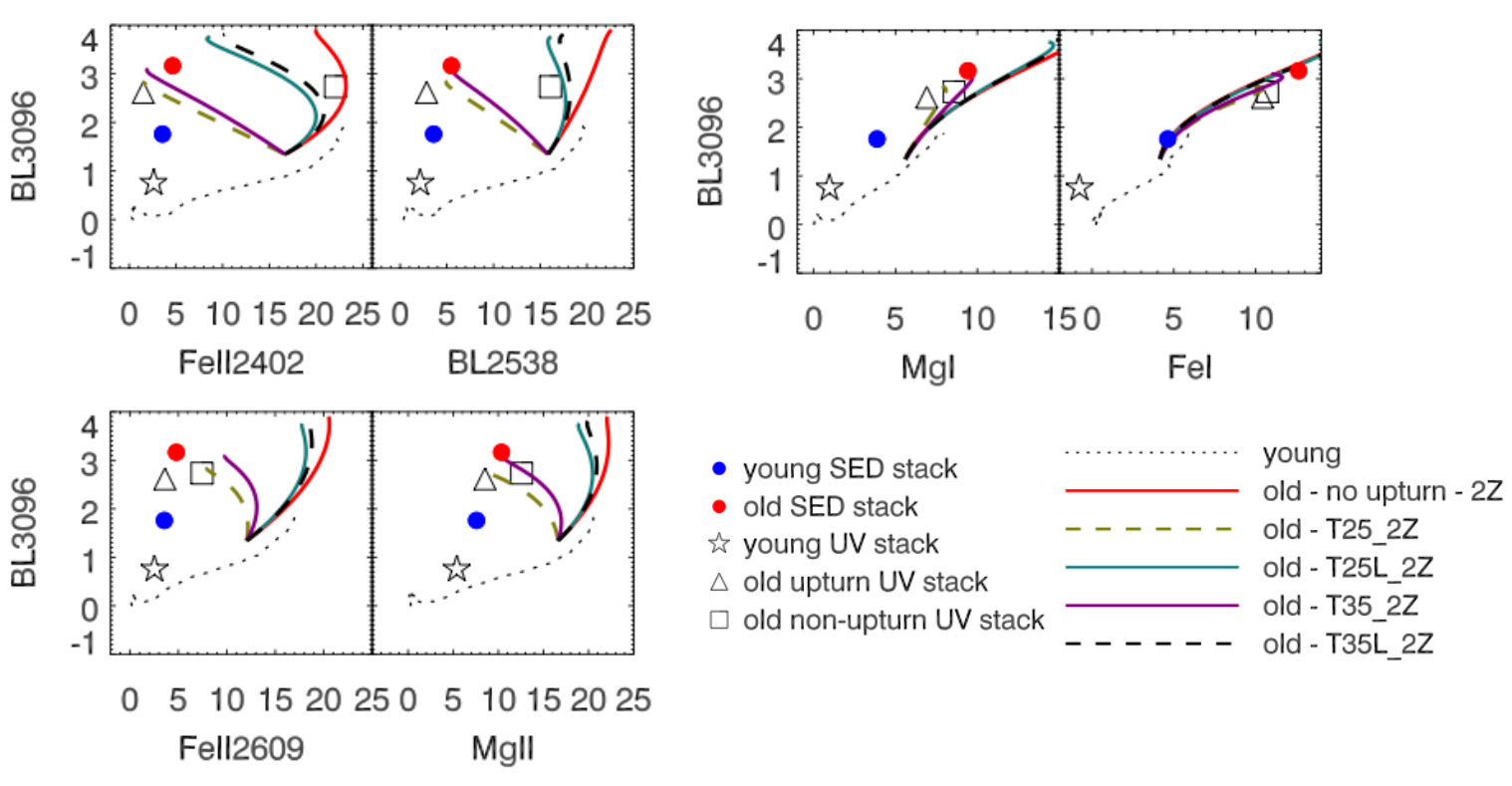}}
\caption{Theoretical model grids. Lines of constant metallicity are shown for various models, see the legend for details. EW values for master stacks with a young UV age, an old UV age with an upturn contribution and old UV age without an upturn are shown by the black star, triangle, and square respectively. The blue and red filled circles show the EWs for young and old optical SED age stacks respectively}
\label{fig:grid3}
\end{figure*}

\section{Quantitative Analysis \& Results}
\label{sec:analysis}

In order to perform the quantitative analysis all model spectra were smoothed to a resolution of 3\AA \ to match that of the BOSS data.\footnote{Note that we have verified that the strength of indices passing from 3\AA\ to the 6\AA\ of the IUE-based empirical models do not change appreciably.} 

On the smoothed spectra the UV ages and metallicities of the UV-bright population in BOSS galaxies were calculated by simultaneously minimising the quadratic distance between the EWs calculated from the data and those of the models for a set of indices using the same software and code as in M09. We classify galaxies with derived UV ages $\le$ 1Gyr as young as this is the end of the young M09 models. This is also the age of the onset of the red giant branch which marks the start of an old population.\footnote{We have tested the effect of varying this split (splitting at both 2 and 3 Gyrs) and find the change in results to be negligible.}

\subsection{Index-index grids in the UV}
Before showing the calculated ages, we visualise data and models in a grid-like form, plotting index vs index, as is usually done for absorption features in the optical. Note that we have further explored the effect of reddening and show it to be negligible for most indices, the exception being Fe I (see Appendix \ref{subsec:dust}). Hence we do not consider reddening in these grids and in the subsequent analysis.

Figure \ref{fig:grid3} shows model grids for 7 of the mid-UV indices; Fe II 2402, BL2538, Fe II 2609, Mg I, Fe I, and BL3096. In these plots lines of constant metallicity are plotted to create a grid like form. The original young theoretical solar metallicity M09 and extended bhb models are shown in Figure \ref{fig:grid3} (a) along with the other solar models; T10, T25, T40, T40L, and the old - no upturn model. Figure \ref{fig:grid3} (b) shows the original young twice solar theoretical M09 model in comparison to the $2Z_{\odot}$ models; T25\_2Z, T25L\_2Z, T35\_2Z, T35L\_2Z, and old - no upturn - 2Z.

Large open symbols show the index values measured on the master stacks. The black star shows the master stack created from galaxies selecting young UV ages; the black triangle, those selecting old ages with an upturn contribution; and the black square, those selecting old ages without an upturn. Also shown are stacks created by splitting the SED age calculated in the Portsmouth galaxy product, found from fitting star forming templates (see Section \ref{subsec:product}). Galaxies with SED ages above 1 Gyr were stacked and the index values are shown by the red circle, those with lower ages are stacked to create the young SED age stack shown by the blue circle.

The spread in values shown mainly corresponds to a spread in age with the EWs increasing as age increases, as seen in Figure \ref{fig:ex_trend}, and choice of upturn model. With this in mind we focus our attention on using these indices to differentiate between the ages of the populations contributing to the UV. It should be noted that metallicity does have some effect on the indices with higher metallicity models having higher EW values.

The young UV stack lies at low EW values, consistent with the younger M09 models shown by the dotted black lines. In the case of the old stacks there is a split between the values seen for those with and without an upturn component, black triangle and square respectively. In the case of the Fe II 2402 and BL2538 vs BL3096 grids a clear split is seen with the old no upturn stacks falling near the old no upturn model shown by the solid red line with the upturn stacks falling towards the left of the grid. Hence this index combination seems very promising in selecting and separating an old population with and without an upturn contribution.

\setcounter{figure}{12}
\begin{figure*}
\includegraphics[width=0.9\textwidth]{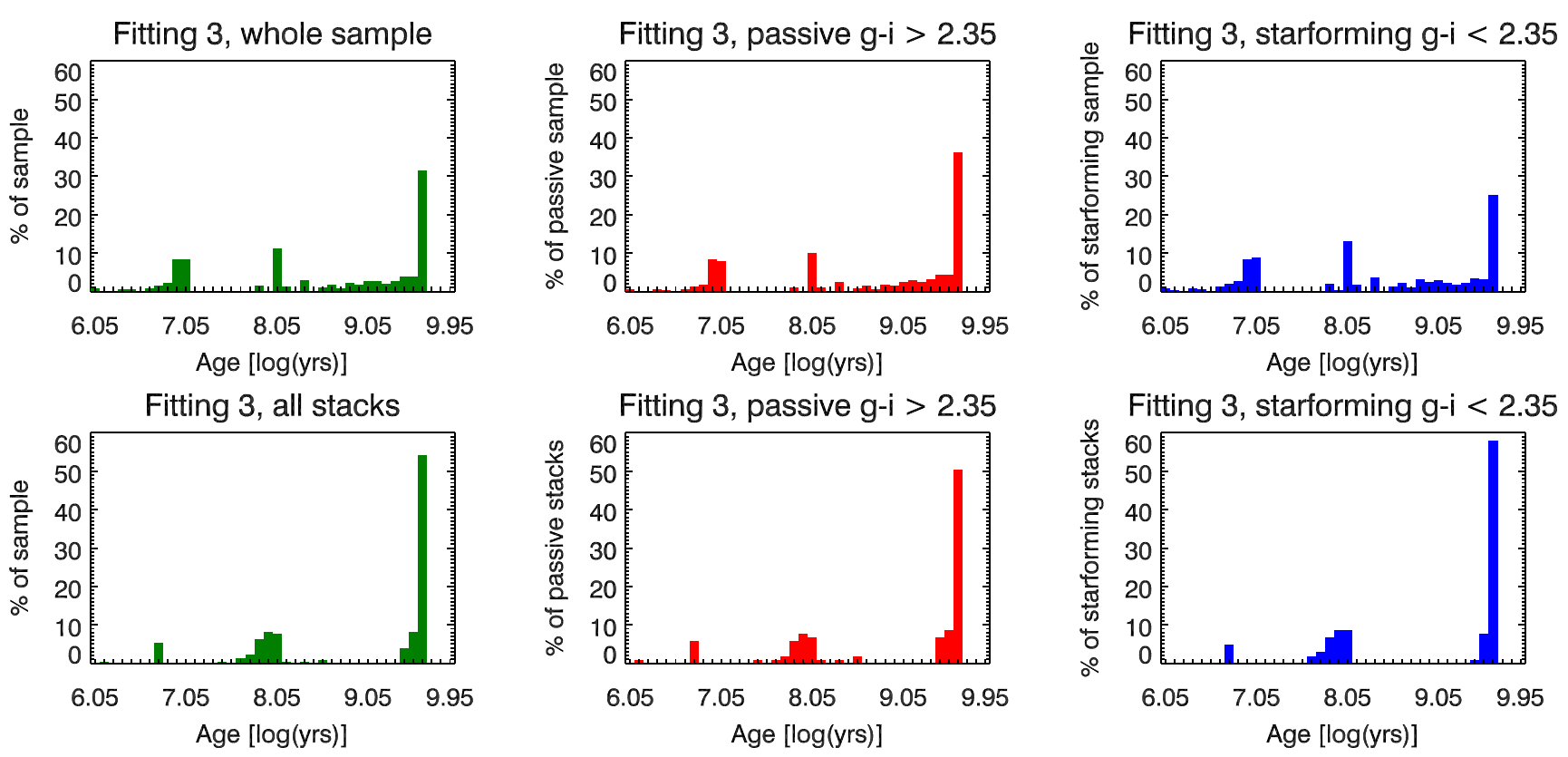}
\caption{Range of UV ages calculated from the fitting of 3 indices shown to be non-degenerate between young and old populations; Mg I, Fe I, and BL3096. All models were fitted simultaneously excluding the no upturn model. \textit{LH} -  entire galaxy sample. \textit{M} - galaxies with $g - i > 2.35$ classified as passive ($\sim 51\%$ of entire sample). \textit{RH} - galaxies with $g - i < 2.35$ classified as star forming ($\sim 49\%$ of entire sample). The relevant stacks for each sample are shown in the lower panel.}
\label{fig:hist_cut}
\end{figure*}

This split becomes less pronounced in the rest of the grids with the old no upturn stacks continuing to show higher EW values than the old upturn stacks, as expected from Figure \ref{fig:ex_trend}. In the case of the Fe II 2609 and Mg II vs BL3096 grids the values of EWs in the old stacks are much lower than those in many of the old models. This is a feature seen in the distribution of EWs seen in Figure \ref{fig:coverage}, with the peak in EWs of stacks seen towards the lower end of the model range. This may be due to an offset of the index in the synthetic stellar spectra (see Section \ref{sec:line} for details). For these two indices the values in the stacks remain towards the lower half of the model range, this could be due to several effects including elemental abundance ratios, contamination from emission lines, and galactic outflows. 
It should be noted that indices tracing Mg and Fe may also be affected by alpha enhancement and abundance ratio effects causing the data to lie away from the models which do not (yet) take account of such effects. As well as this we have shown the Fe I index to be the most affected by the presence of interstellar dust, see Appendix \ref{subsec:dust}.

In the case of the $Z_{\odot}$ upturn models, the cooler upturn models favour the EW region surrounding the upturn stack  (black triangle) in most of the model grids, with the T25 model with a temperature of 25,000 K falling more closely to the data in the majority of panels. The hottest and lowest fuel model, T40L, lies much closer to the old stack without an upturn contribution, which may simply reflect that this hot model peaks at wavelengths much shorter than those covered by these mid-UV indices and simply cannot be constrained by these data.  

For the $2Z_{\odot}$ case, models with lower fuel follow the trends of the higher fuel models with differing temperatures (the T25L\_2Z model following the T35\_2Z model and the T35L\_2Z model the T25\_2Z). Similarly to the $Z_{\odot}$ model the lower temperature and higher fuel model lies more closely to the upturn stack.

The SED age stacks show an interesting behaviour with the young optical SED stack (blue circle) falling between the young UV (black star) and old UV upturn stack (black triangle), and the old optical SED stack (red circle) falling between the old UV upturn and old no UV upturn stack (black square). The proximity of the EW values of the young optical SED to those of the old UV upturn stack confirm that indeed the UV contribution may be fit equally well or even better (see next section) by an old hot population rather than residual star formation.

\subsection{UV ages for different models.}

By focusing on the 3 non-degenerate indices shown to be able to distinguish between a young and an old UV-bright population (Mg I, Fe I, and BL3096), and fitting both the original theoretical and empirical M09 models and all the theoretical models with a contribution from old UV-bright stars discussed in Section \ref{model:th}, we obtain the distribution of UV ages shown in Figure \ref{fig:hist_cut}. We do not fit the old models without an upturn at this point as the indices used lack the abilty to distinguish between the two types of old population (with and without an upturn contribution) on their own.

Approximately 48\% of our working sample is best fitted by a model with a contribution from an old UV-bright population, with more passive galaxies selecting those ages than star forming ones. In the case of the stacks, all stacks created with galaxies selecting ages less than 1 Gyr continue to select young UV populations as their best fit, with the majority of those stacked from older ages still selecting ages greater than 1 Gyr\footnote{We note that there are 3 distinguished peaks that appear in these distributions. Further interpolation of the models to produce finer grids in age dilutes the distribution slightly towards older ages but the peaks remain pronounced. This suggests the origin may be intrinsic to the galaxies themselves, an interesting result to be explored in the future.}.

To examine how well our modelling fits the data we calculated the reduced $\chi^2$ for each of the best fits found. This is derived by calculating the distance between the index strengths in the data and its best fit model and dividing by the number of degrees of freedom (the number of indices fitted). The distribution of these values can be seen in Figure \ref{fig:chi_ex}, with the values for the individual spectra shown by the black hashed histograms and the young and old stacks by the solid blue and red respectively. The top panel shows the theoretical models and the bottom the empirical.

 \setcounter{figure}{13}
\begin{figure}
\includegraphics[width=0.45\textwidth]{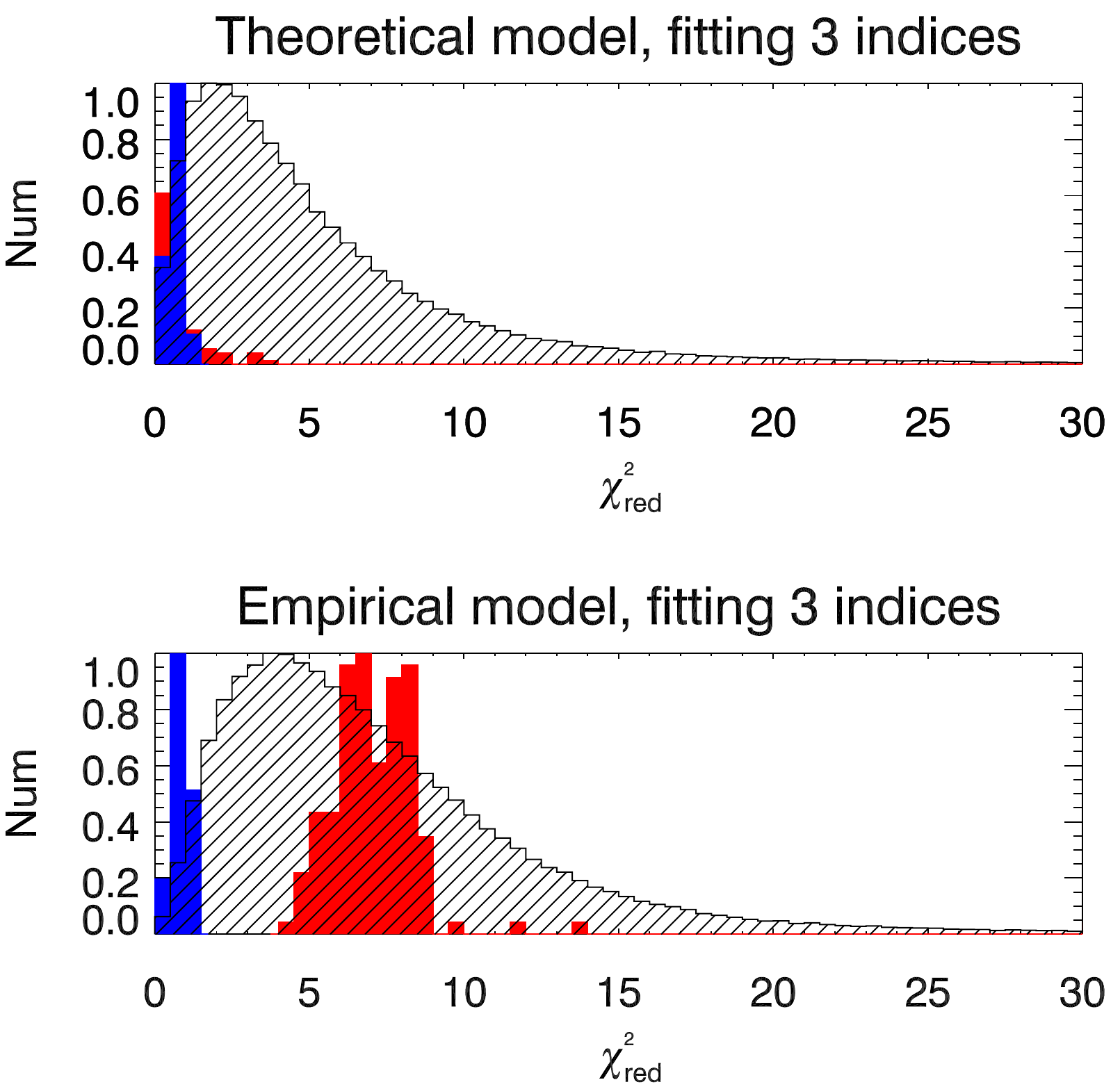}
\caption{Distribution of reduced $\chi^2$ values for the best fits found fitting the 3 non-degenerate indices (Mg I, Fe I, and BL3096) with the individual spectra shown by the black hashed histograms and the young and old UV age stacks by the solid blue and red histograms respectively. \textit{Top panel -} shows the values when simultaneously fitting the M09, bhb, and suite of PAGB theoretical models. \textit{Bottom panel -} shows the values when fitting the M09 empirical model. Each distribution has been normalised to 1.}
\label{fig:chi_ex}
\end{figure}

\setcounter{table}{2}
\begin{table}
\caption{A breakdown of the number of galaxies selecting different UV models. The favoured upturn models are highlighted in bold.}
\begin{center}
\begin{tabular}{ | c | c | c | }
Model & Num  & \% of entire sample \\ \hline
\hline
young & 142,063 & 51.7 \\
\textbf{old - T10} & \textbf{33,295} & \textbf{12.1} \\
old - T25 & 3,856 & 1.4 \\
\textbf{old - T25\_2Z} & \textbf{37,718} & \textbf{13.7} \\
old - T35\_2Z & 14,648 & 5.3 \\
old - T40 & 2,259 & 0.8 \\
old - bhb & 2,622 &  1.0 \\
old - T25L\_2Z & 5,100 & 1.9 \\
old - T40L & 891 & 0.3 \\
old - T35L\_2Z & 3,686 & 1.4 \\
old - no upturn - 2Z & 16,488 & 6.0 \\
old - no upturn & 12,035 & 4.4 \\
\label{tab:choice}
\end{tabular}
\end{center}
\end{table}

\setcounter{table}{3}
\begin{table}
\caption{Average properties of galaxies selecting different UV models The favoured upturn models are highlighted in bold..}
\begin{center}
\begin{tabular}{ | c | c | c | c | }
Model & $< z >$  & $<$ Log(Mass) $>$ & $< g - i >$ \\ \hline
\hline
young & 0.671 & 11.47 & 2.37 \\
\textbf{old - T10} & \textbf{0.663} & \textbf{11.55} & \textbf{2.50} \\
old - T25 & 0.664 & 11.59 & 2.58 \\
\textbf{old - T25\_2Z} & \textbf{0.663} & \textbf{11.59} & \textbf{2.59} \\
old - T35\_2Z & 0.661 & 11.66 & 2.71 \\
old - T40 & 0.660 & 11.66 & 2.71 \\
old - bhb & 0.656 & 11.66 & 2.71 \\
old - T25L\_2Z & 0.657 & 11.67 & 2.75 \\
old - T40L & 0.658 & 11.69 & 2.75 \\
old - T35L\_2Z & 0.661 & 11.67 & 2.75 \\
old - no upturn - 2Z & 0.655 & 11.68 & 2.76 \\
old - no upturn & 0.655 & 11.68 & 2.77 \\
\label{tab:props}
\end{tabular}
\end{center}
\end{table}

The theoretical models fit the data more closely with a lower average $\chi^2$ than the empirical, 8.92 and 10.95, with the distributions peaking around 2 and 4 respectively. In the case of the stacks, the theoretical models fits more closely with an average value of 0.95 and little spread. The empirical model shows a bimodal distribution with the young stacks being fit well, shown by the peak in the blue histogram around 1. The older stacks are fit less well, shown by the second peak in red above 5, due to the limited range of the empirical model in both age and EW. For these reasons, and because of the possibility of investigating a variety of UV upturn models, from now on we proceed with the theoretical models only.

As shown in Figure \ref{fig:ex_trend}, models containing old stellar populations, both with and without a contribution from old UV-bright stars, show an increase in index strength for the 3 non-degenerate indices used in our age calculation. To determine whether the old UV ages derived from our fitting are due specifically to an old UV-bright population, we forced the objects determined to have an old UV age from fitting 3 indices to pick between the old models with and without an upturn component. We add in the 4 other indices (Fe II 2402, BL2538, Fe II 2609, and Mg II) for this calculation as above 3 Gyrs they show the ability to distinguish between an old component with and without an upturn contribution.

It was found that $78\%$ of the galaxies selecting old UV ages preferred models with an old UV-bright population present. This equates to $\sim37\%$ of our entire working sample of BOSS galaxies, which we define as our upturn confirmed sample. This percentage is significantly higher than the results of \citet{Yi:2011} who find that only $5\%$ of cluster elliptical galaxies show a UV upturn using a new UV classification scheme based on far-UV, near-UV, and optical photometry from GALEX and SDSS. It should be noted however that a further 68\% of their sample are classified as "UV-weak", those without residual star formation but also without a strong upturn slope. Similar galaxies in our sample may select an upturn SSP due to the strength of the indices alone. We could in the future check this with rest-frame UV spectroscopy to study the shape of any upturn present, as well as photometry to test our galaxies against the \citet{Yi:2011} classification scheme.

Table \ref{tab:choice} shows a breakdown of the number of galaxies selecting each of the different UV models. Of the models containing an upturn contribution: the T10 and T25\_2Z models are preferred, with 12.1\% and 13.7\% of the working sample selecting each respectively; this points towards the ideal temperature range of stars contributing to the UV upturn being between 10,000 - 25,000K, with higher metallicity models needing higher temperatures. This is as expected from the model grids in Figure \ref{fig:grid3} and is consistent with the temperature required to model local UV upturn galaxies. It should also be noted that models with higher fuels are preferred over those with lower values, hinting towards an ideal fuel of $6.5\cdot10^{-2}~\textnormal{M}_{\odot}$.

\

The average properties of the galaxies selecting each model can be seen in Table \ref{tab:props}. On average the least massive galaxies select young ages without an upturn component and have significantly bluer $g - i$ colours. Galaxies selecting upturn models fall between the young and old no upturn  populations with the temperature and metallicity of the model being the driving parameter. $g - i$ colours increase and become redder as the upturn temperature increases with higher metallicity models having redder colours than those with solar metallicity. There is a slight trend seen in the average redshift and mass for the galaxies selecting each model. However, due to the small variation in values, this is not seen as significant.

\subsection{UV Diagnostics vs Broadband Spectral Fitting}
\label{subsec:uv_sed}

\setcounter{figure}{14}
\begin{figure*}
\includegraphics[width=0.9\textwidth]{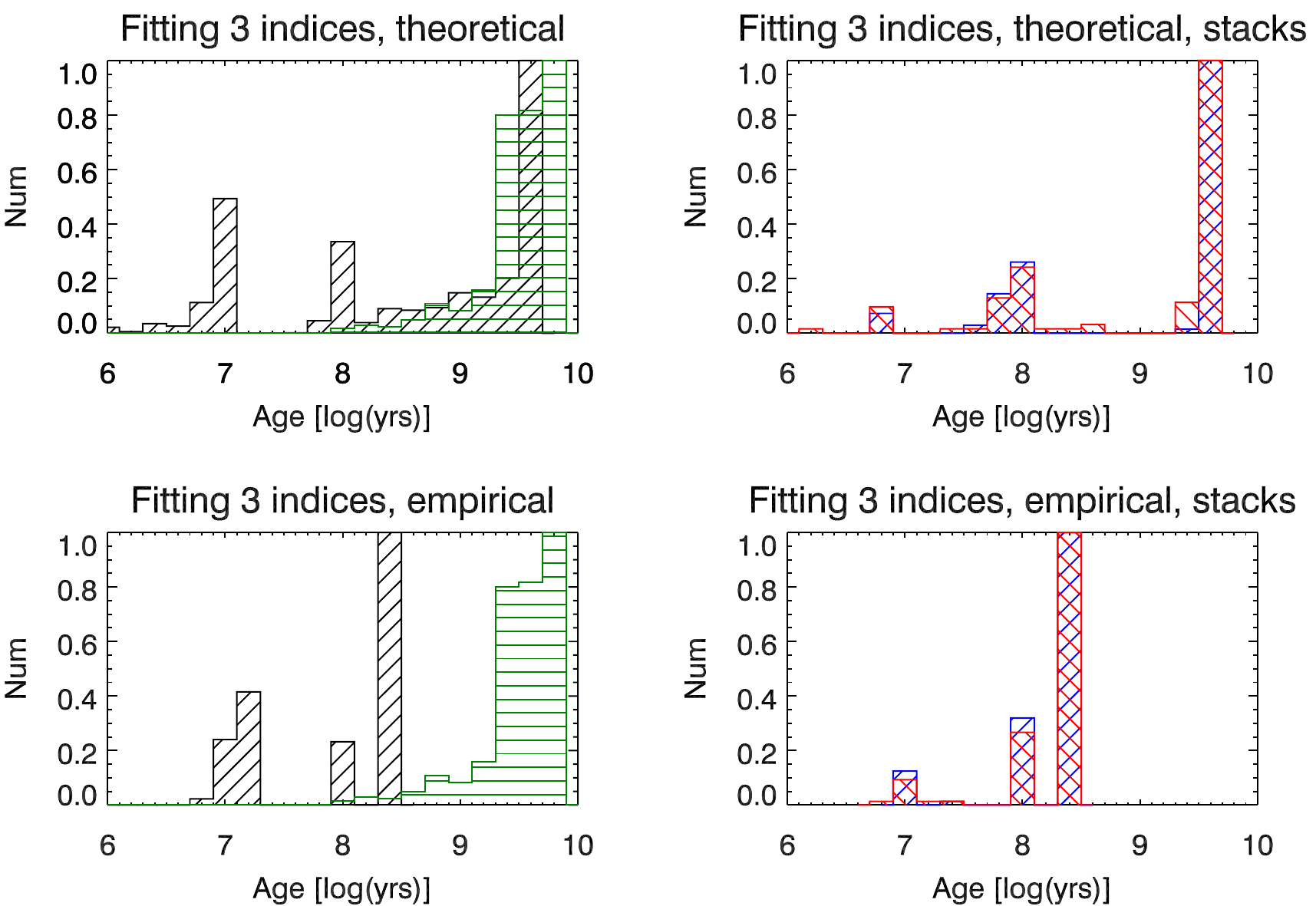}
\caption{Comparison of the distribution of ages derived from full SED fitting from $u$ to $z$, shown in green, and from the 3 non-degenerate indices, black, for individual galaxies. The ages of the stacks, with an average $g - i <$ 2.35 are shown in blue with those above in red. The top panels show the ages found from fitting the extended theoretical model, with the bottom panels showing those from fitting the empirical. Each distribution has been normalised to 1.}
\label{fig:hist_star}
\end{figure*}

Using the age calculated from the UV absorption indices and those calculated from the best fit broad-band SED it is possible to quantify the difference in physical parameters as extracted from different spectral regions.

The SED age was selected based on the colour of the galaxy, those with $g - i < 2.35$ were classified as star forming and as such have the age calculated by the star forming stellar population model in the BOSS Portsmouth galaxy product described in Section \ref{subsec:product}. Those with $g - i \ge 2.35$ have that calculated from the passive model.

Figure \ref{fig:hist_star} shows the distribution of UV ages in comparison to those found from full SED fitting. The top panels show the theoretical models with the empirical model shown in the bottom panels. In both cases the UV ages for individual galaxies are shown by the black hashed histograms and the SED age by the green in the lefthand panels. The UV ages derived for the stacks using each model set are shown in the righthand panels, with stacks with an average $g - i < 2.35$ being shown in blue and those above in red.

When fitting the empirical model there is a bias towards the oldest age present in the model, log(age [yrs]) = 8.3, for both the individual galaxies and the stacks. This age is far below those found from SED fitting, an offset of log(age [yrs]) = 1.71 on average, but this is a restriction imposed by the nature of the model. In the case of the theoretical models the ages calculated from the UV indices extend to higher ages showing a peak similar to those found from full broadband SED fitting. These ages agree more with a lower average offset of log(age [yrs]) = 1.16. 

\setcounter{figure}{15}
\begin{figure}
\centerline{\includegraphics[width=0.36\textwidth]{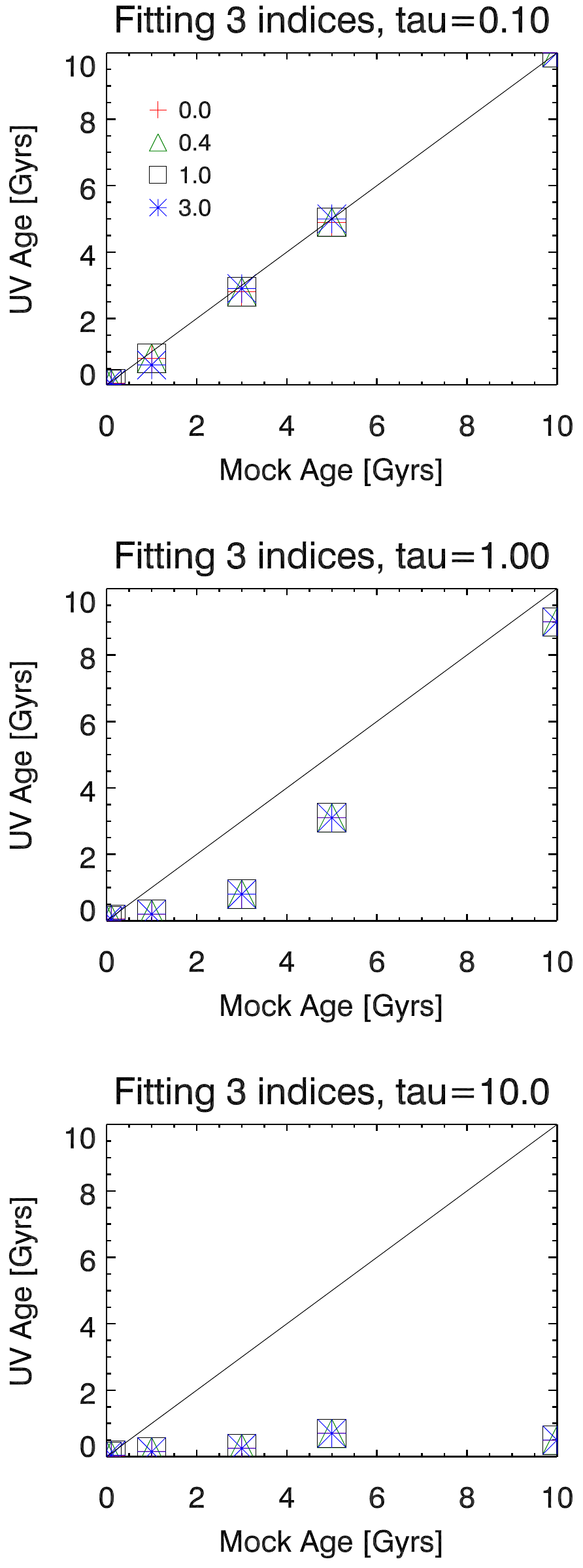}}
\caption{Meaning of the UV age. The mock age denotes the age of the oldest population present in composite models not including any old UV component. The ages derived from fitting Mg I, FeI, and BL3096 can be seen for the three different star formation rates (star formation timescales; $\tau$ = 0.10, 1.00, 10.0 Gyrs). The different symbols denote the different dust attenuation values shown in the first panel. The solid black line donates a one to one relationship.}
\label{fig:csp}
\end{figure}

The peaks seen towards older ages for both the UV and SED ages confirm the result that BOSS galaxies are predominantly old (\citet{Maraston:2013boss}).

More quantitatively, we have calculated the average broad-band SED age for each of the UV-age master stacks by averaging the star forming SED ages found in the Portsmouth galaxy product (Section \ref{subsec:product}) for each of the galaxies within each stack. The young UV stack was found to have an average SED age of 3.6 Gyrs, hence younger than the two old stacks, 4.1 and 4.4 Gyrs for the old upturn and old no upturn stacks respectively. The SED age of the young UV stack is much higher than that from fitting the UV indices due to the different populations dominating the different fitting regions. However this SED age being lower than those of the two old UV stacks supports our use of the mid-UV indices to differentiate the ages of stellar populations. The similar ages of the old upturn and old no upturn stacks support the existence of old hot populations contributing to the UV, with both producing older SED ages averaged over the $ugriz$ bands but showing differences in the details of the UV spectrum.

\section{Fitting composite populations}
\label{model:ex}

We now proceed by investigating the physical properties of the galaxies best-fitted by an old UV-bright model.
So far we have employed single-bursts as fitting units. However, real galaxies are made of stellar generations, and it is always interesting to understand the comparison between a single-burst approximation age and the one from a composite model fit. Moreover, the second one will allow us to quantify the mass fractions pertaining to our putative old UV upturn populations.

\subsection{UV-derived vs true ages}
\label{subsec:comp}

To investigate the meaning of a UV-derived age we use mock galaxies whose ages are defined as input and calculate their UV ages using the same method we use for the data, namely fitting various UV spectral indices. The mocks are generated using the \citet{Maraston:2005} software and comprise solar metallicity Composite Stellar Population (CSPs) models with exponentially declining star formation rates (with star formation timescales; $\tau$ = 0.10, 1.00, and 10.0 Gyrs) and various ages for oldest population present ($t$ = 0.1, 1.0, 3.0, 5.0, and 10.0 Gyrs). Models are calculated for four different dust attenuation values ($a$ = 0.0, 0.4, 1.0, and 3.0) and smoothed to match the resolution of our models. Figure \ref{fig:csp} compares the UV ages calculated for each CSP fitting the Mg I, Fe I, and BL3096 indices ($y$-axis) with the entry age of each model, using CSPs not including any UV from old populations. 

For composites with high star formation rates ($\tau$ = 10.0 Gyrs) the UV-derived age is always young. In this case the UV light is dominated by the young, newly formed stars and as such is the population traced by our fitting method. As the star formation drops ($\tau$ = 1.00 Gyrs) the UV age becomes older and traces the age of the oldest population present remarkably well. This is also true for the composite with lowest star formation rate ($\tau$ = 0.10 Gyrs). 

It should be noted that the amount of dust present in the mock galaxies does not affect the UV age calculated when fitting multiple indices. This confirms our theoretical investigation of the effect of dust on indices, Appendix \ref{subsec:dust}. This property will be important when using the indices for fitting dusty, star-forming galaxies.

\setcounter{figure}{16}
\begin{figure}
\centerline{\includegraphics[width=0.48\textwidth]{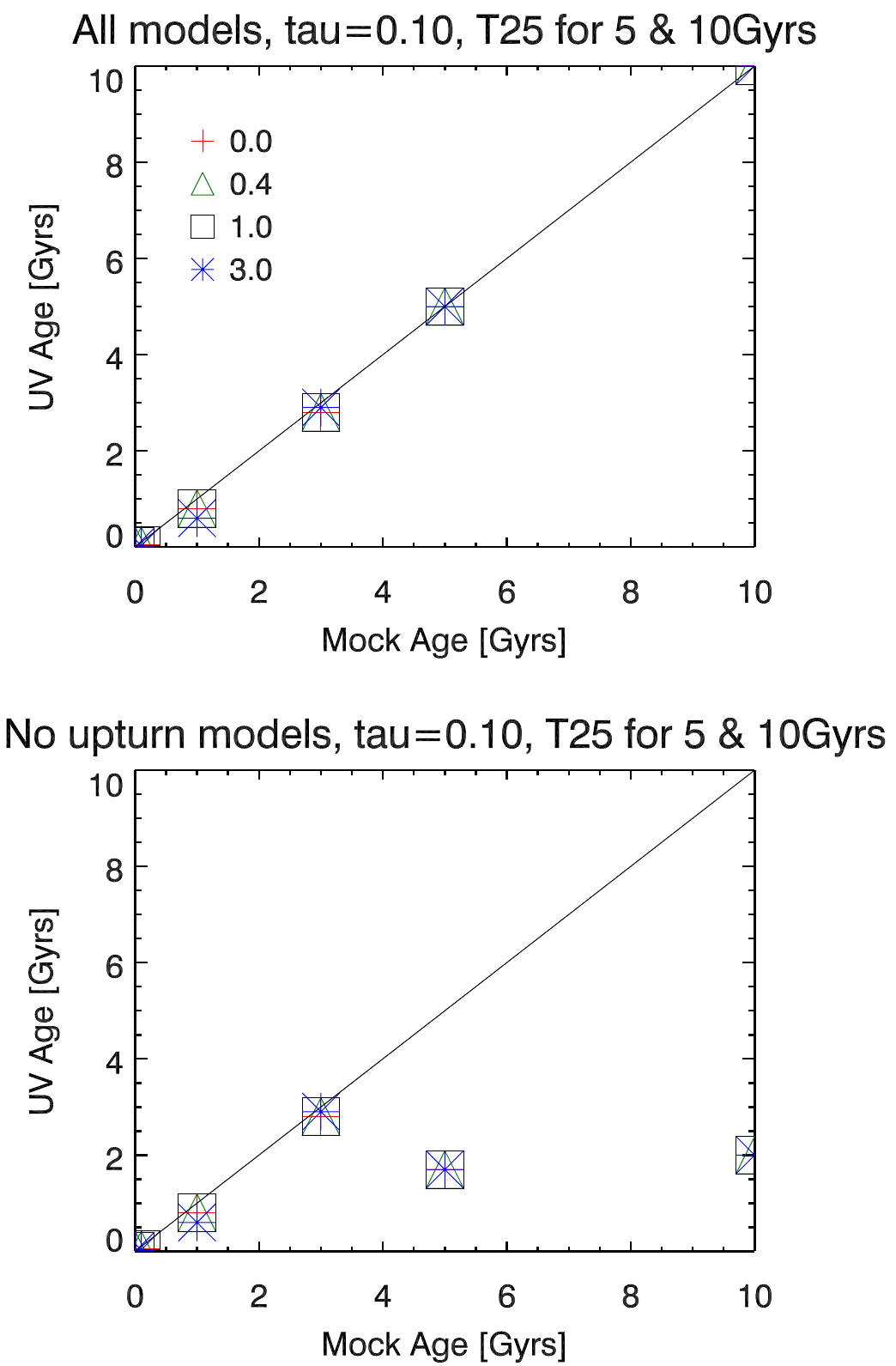}}
\caption{As in Figure 15, but here the mocks include a 50\% contribution from the upturn T25 model at ages 5 \& 10 Gyrs. When fitting this composite mock with all models including those with an upturn (top panel) the actual age is well recovered. When instead the fitting models do not include the upturn, the recovered age is underestimated (bottom panel).}
\label{fig:hybrid}
\end{figure}

With our models we can now test what happens when a (mock) galaxy contains a fraction of upturn population and gets fitted with models including such an effect or not.

To this end we create hybrid spectra at ages 5 \& 10 Gyrs using a 50\% contribution from the $\tau$ = 0.10 Gyrs CSP and 50\% from the T25 model of the corresponding age. We then fit the hybrids and the original younger CSPs, using all our models, as shown in the top panel of Figure \ref{fig:hybrid}, and again using only models without an upturn contribution in the bottom panel. When fitting all models we recover the ages well across all ages with the oldest two, the CSP/upturn hybrids, selecting ages from the T25 model and the 3 Gyr CSP an age from the old no upturn model. 

However when omitting models with an upturn contribution we severely underestimate the ages of the oldest mocks, with their upturn contributions forcing the selection of younger ages. This shows that when fitting features in the UV region it is crucial to include contributions from the UV upturn in order to avoid systematically younger ages due to model restrictions.

As our BOSS galaxies do not feature strong star formation (see \citet{Thomas:2013boss}),  our testing using mock composite populations confirms that our SSP results are robust and therefore we proceed investigating further properties.

\setcounter{figure}{17}
\begin{figure*}
\includegraphics[width=0.9\textwidth]{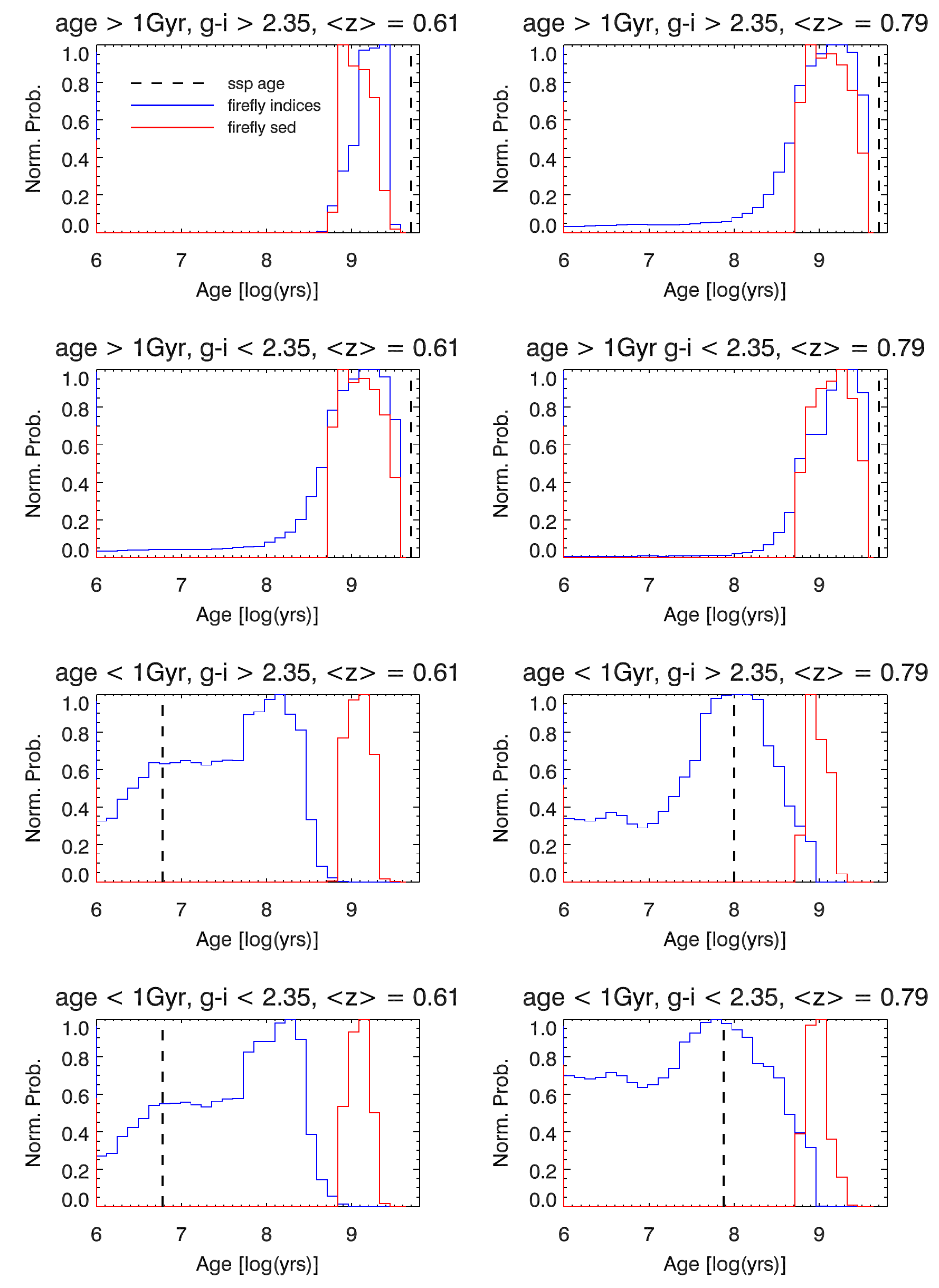}
\caption{Probability distribution of UV ages for 8 stacks found using {\sc firefly} to fit Mg I, Fe I, and BL3096, in blue, with those found from fitting the spectral region between 2250 - 3200\AA \ in red. The dashed black line shows the SSP age from fitting 3 indices using the $\chi^2$ minimisation.The distributions have been normalised to 1. The mass fraction of the populations present can be derived from the area under each histogram.}
\label{fig:pdfs3_stack}
\end{figure*}

\subsection{Mass fractions from UV-bright populations.}
\label{subsec:mass_frac}

A crucial piece of information is how much galaxy mass is involved in the UV upturn. We are able to investigate the possibility of multiple populations contributing to the UV spectra of our BOSS sample, and their relative mass fractions, by adapting a new full spectral fitting code, {\sc firefly}, to fit the EWs of the mid-UV indices.

{\sc firefly} was designed to map out inherent spectral degeneracies, work well at low SNR, and allow comparisons between different input stellar libraries, while making as few assumptions about the star formation histories derived as possible by modelling star formation histories as a linear combination of bursts. The {\sc firefly} code is discussed in detail, with tests on mock galaxies and example applications to SDSS objects, in a separate work (\citet{Wilkinson:2015}). 

The probability distribution for stellar ages found from {\sc firefly's} full spectral fitting allows us to give a more comprehensive comparison of the UV and SED properties of our BOSS galaxies.

In this work we use a modified version of {\sc firefly} to fit the Mg I, Fe I, and BL3096 indices to the theoretical models to derive a probability distribution of ages for each individual galaxy and stack. We also fit the mid-UV spectral region between 2250 - 3200\AA \

Figure \ref{fig:pdfs3_stack} shows the age distribution found from {\sc firefly} for a selection of 8 stacks, with varying properties. The age distribution found from fitting the indices, shown in blue, shows relative agreement with the SSP age derived previously, shown by the dashed black lines, with the majority of SSP ages falling near the peak of the probability distributions. In the case of the 4 stacks created from galaxies selecting old UV ages, those in the upper 4 panels, the distribution found from fitting the mid-UV region, shown in red, agrees reasonably well with that from the indices. This is not the case for stacks created using galaxies selecting younger UV ages. The ages found from fitting the mid-UV region are highly discrepant to those found from the other 2 methods as they favour much older ages. This can be explained by the similarities in the shapes of the young and old stacks which all feature a systematic increase in flux towards the bluer end. Therefore the fitting of the UV spectral region cannot be considered reliable.

The stellar mass fraction contributing to these UV ages can be calculated from the area under the unnormalised probability distribution of ages, for both individual galaxies and stacks. For the mass contributing to the upturn we calculate the area in the distribution above 1 Gyr as a fraction of the whole. The distribution of mass fractions for the older populations in our upturn confirmed sample, individual galaxies that select an old SSP UV age with an upturn component over one without, can be seen in Figure \ref{fig:mass_frac}.

\setcounter{figure}{18}
\begin{figure*}
\includegraphics[width=0.85\textwidth]{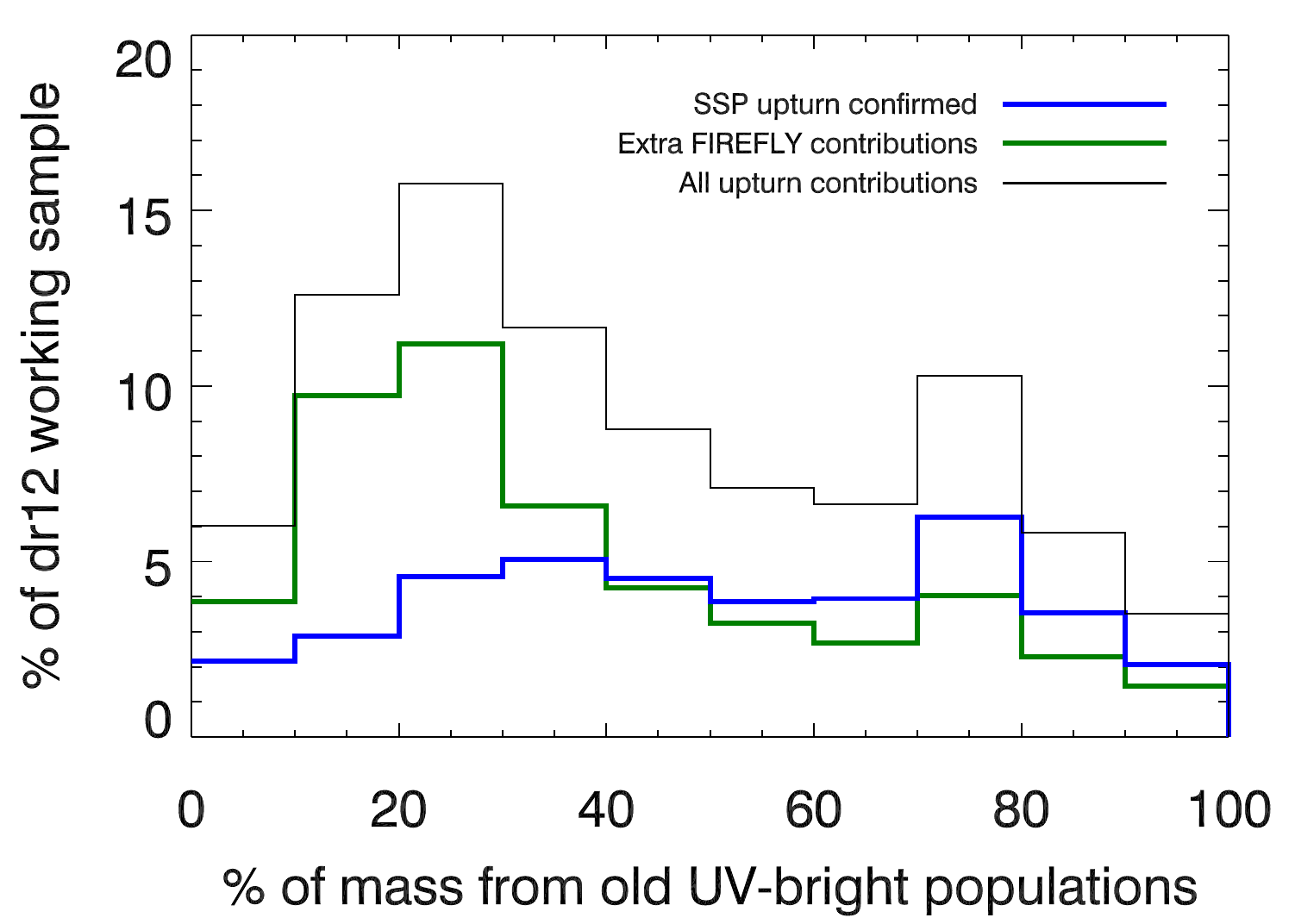}
\caption{The distribution of the stellar mass fraction contributing to UV ages above 1 Gyr. The blue line shows the percentage of galaxies within our working sample with an upturn confirmed SSP age, with various mass contributions from the old component. The green line shows extra galaxies, without an old upturn age from SSP fitting, found to have an old component from the {\sc firefly} composite fit as a percentage of the entire DR12 working sample. The black line shows the sum of these two samples.}
\label{fig:mass_frac}
\end{figure*}

\setcounter{figure}{19}
\begin{figure*}
\includegraphics[width=0.9\textwidth]{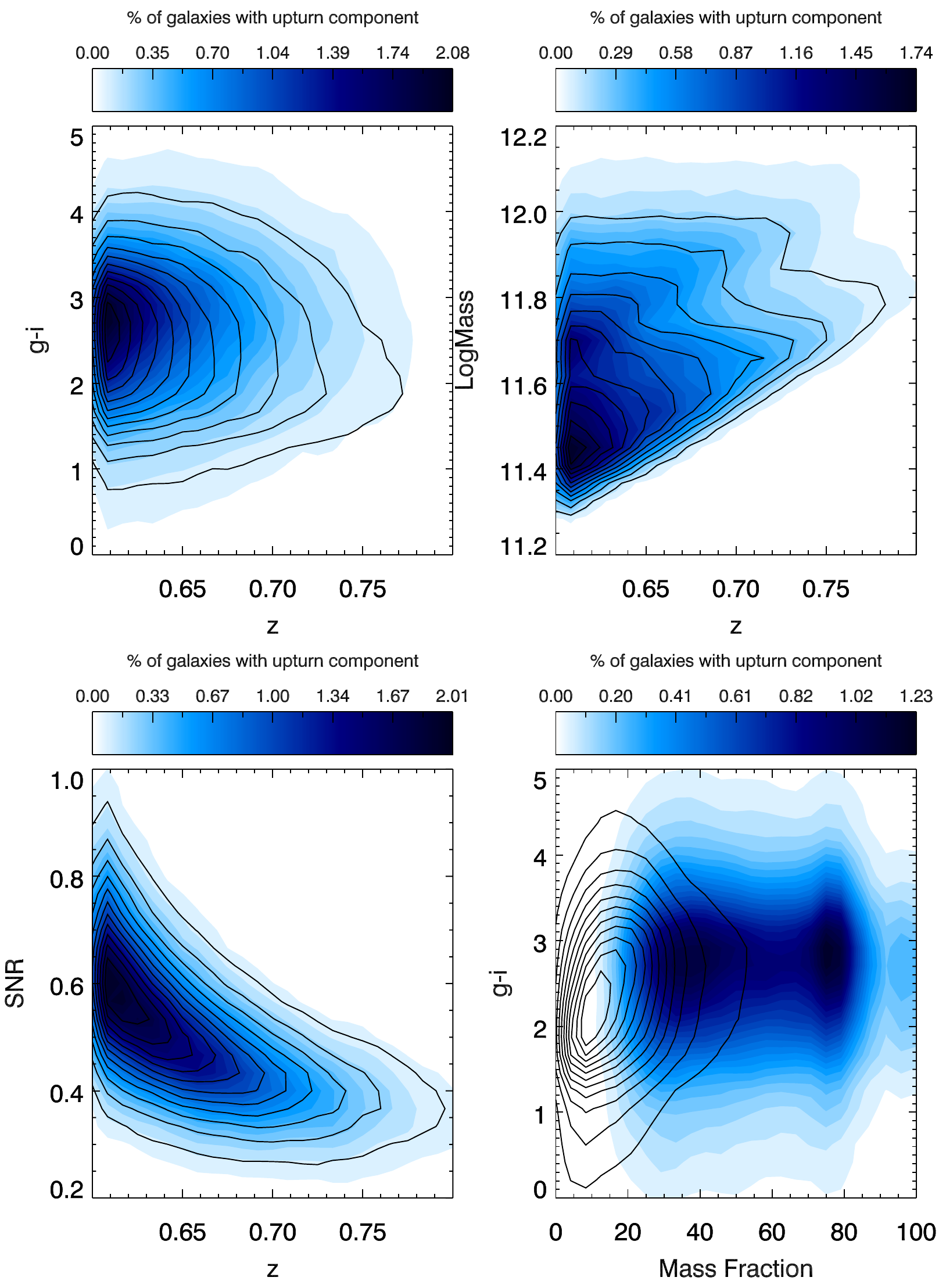}
\caption{The difference in various properties of the SSP upturn confirmed galaxies and the additional galaxies found to have an upturn component from {\sc firefly}. The blue coloured contours show the upturn confirmed sample and the black contours the additional upturn galaxies as a percentage of all galaxies with upturn contributions. Both sets of contours show the number of galaxies as a percentage of all galaxies with upturn contributions.}
\label{fig:upturn_extra}
\end{figure*}

\setcounter{figure}{20}
\begin{figure}
\centerline{\includegraphics[width=0.47\textwidth]{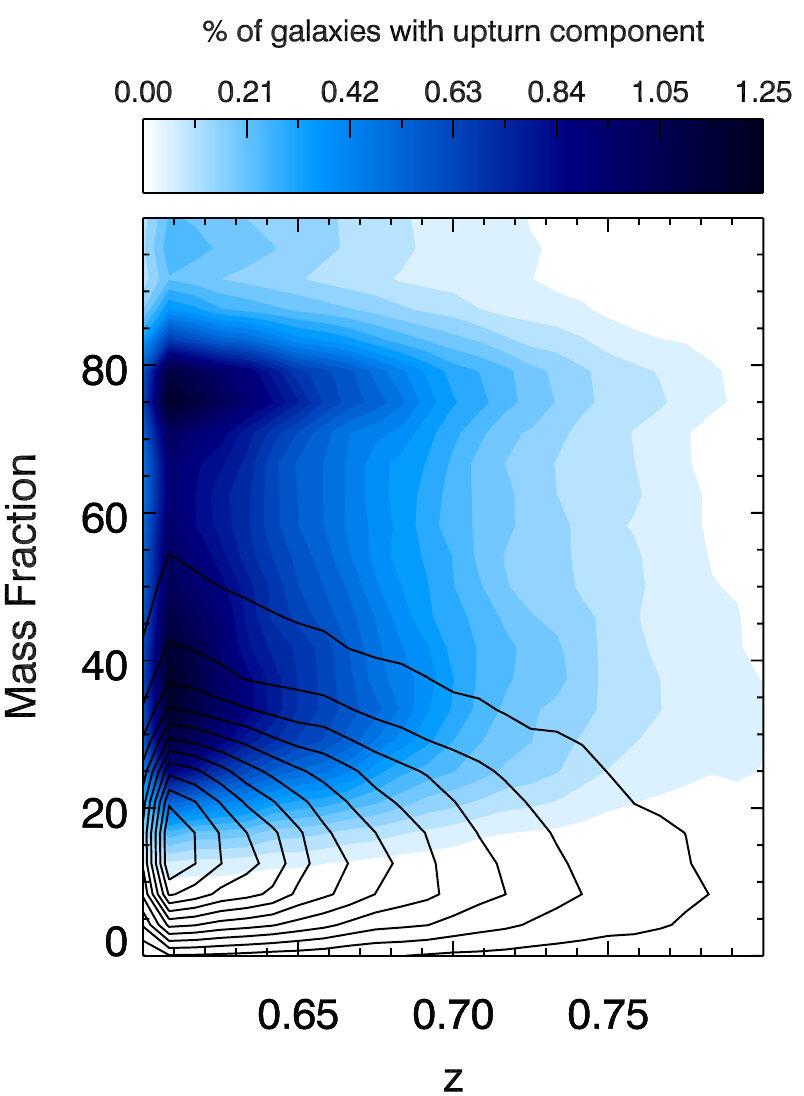}}
\caption{The evolution of mass fractions with redshift for upturn confirmed galaxies, shown by the blue coloured contours, and the additional galaxies found to have an upturn component from {\sc firefly}, in black. Both sets of contours show the number of galaxies as a percentage of all galaxies with upturn contributions.}
\label{fig:upturn_extra_z}
\end{figure}

The distribution in mass fraction of our upturn confirmed sample, shown by the solid blue line, spans a wide range, with at least 10\% of the mass of each galaxy contributing to the UV upturn. The green line shows the extra galaxies within the DR12 working sample that are not part of the upturn confirmed sample but preferentially select an old age component with an upturn contribution over one without when using {\sc firefly} to fit composite populations. It is clear that there are far more galaxies with low mass fractions from old, hot components that we do not classify as upturn confirmed from our SSP fitting. Including all these extra classifications increases the percentage of DR12 galaxies in our working sample with upturn contributions from $\sim$37\% to $\sim$92\%.

\setcounter{figure}{21}
\begin{figure}
\includegraphics[width=0.47\textwidth]{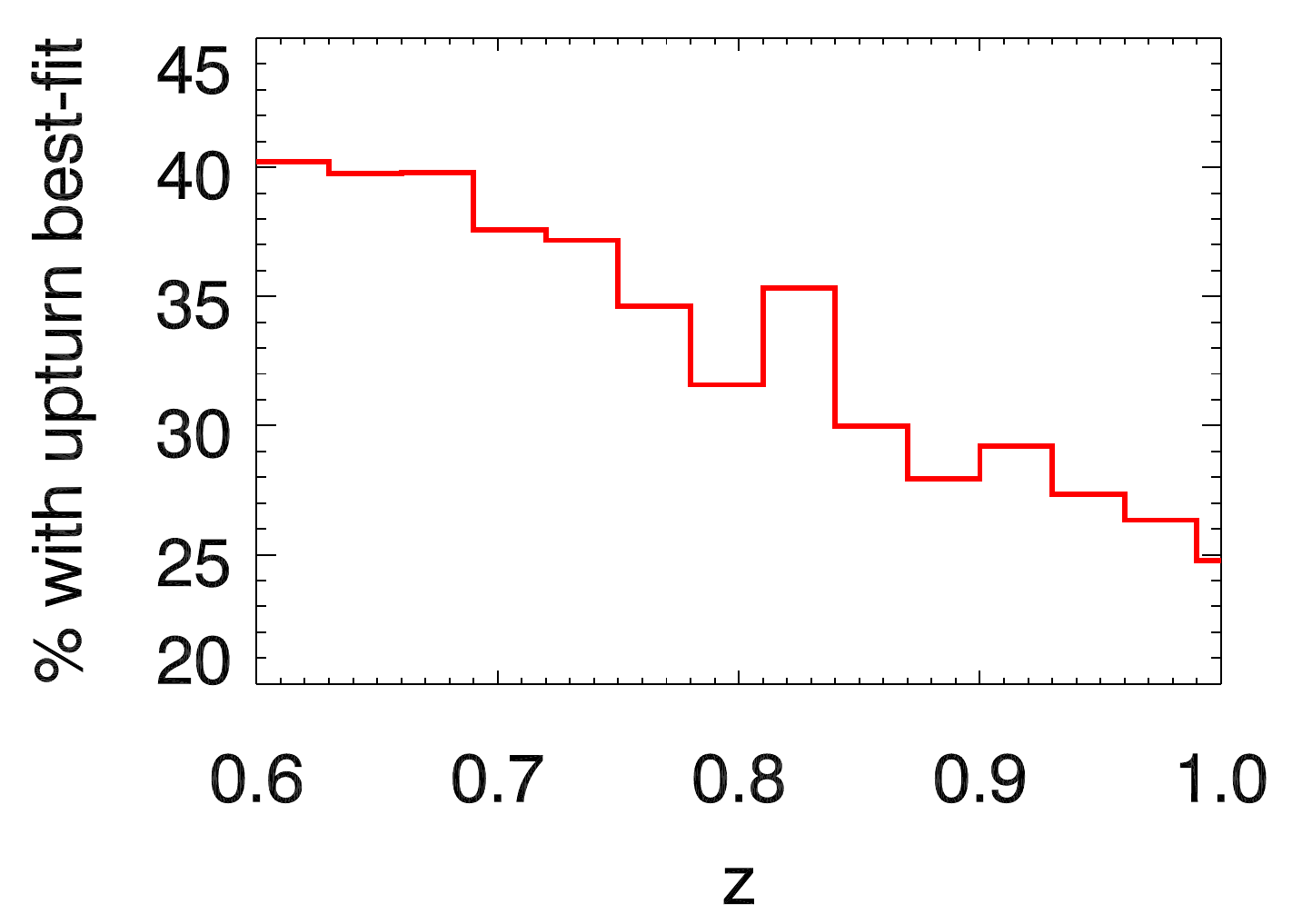}
\caption{Fraction of galaxies within each redshift bin that select an SSP with an old UV-bright population as the best fit solution. Redshift bins start at z = 0.6, increasing in increments of 0.03 up until z = 1.0.}
\label{fig:age_frac}
\end{figure}

\setcounter{figure}{22}
\begin{figure}
\includegraphics[width=0.46\textwidth]{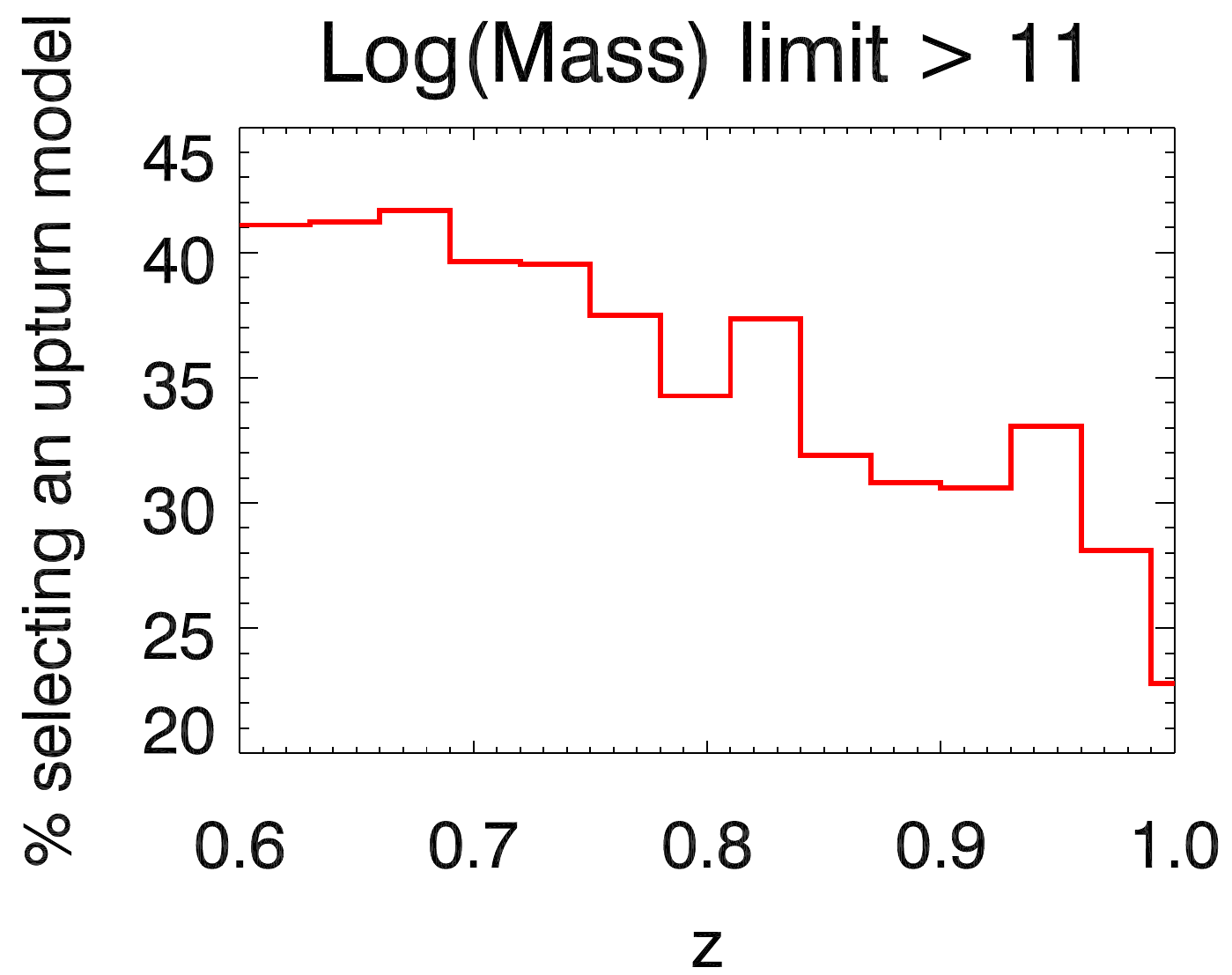}
\caption{As in Figure \ref{fig:age_frac} but with a mass limit such that Log(Mass) $\ge$ 11.}
\label{fig:age_frac_mass}
\end{figure}

\setcounter{figure}{23}
\begin{figure}
\includegraphics[width=0.48\textwidth]{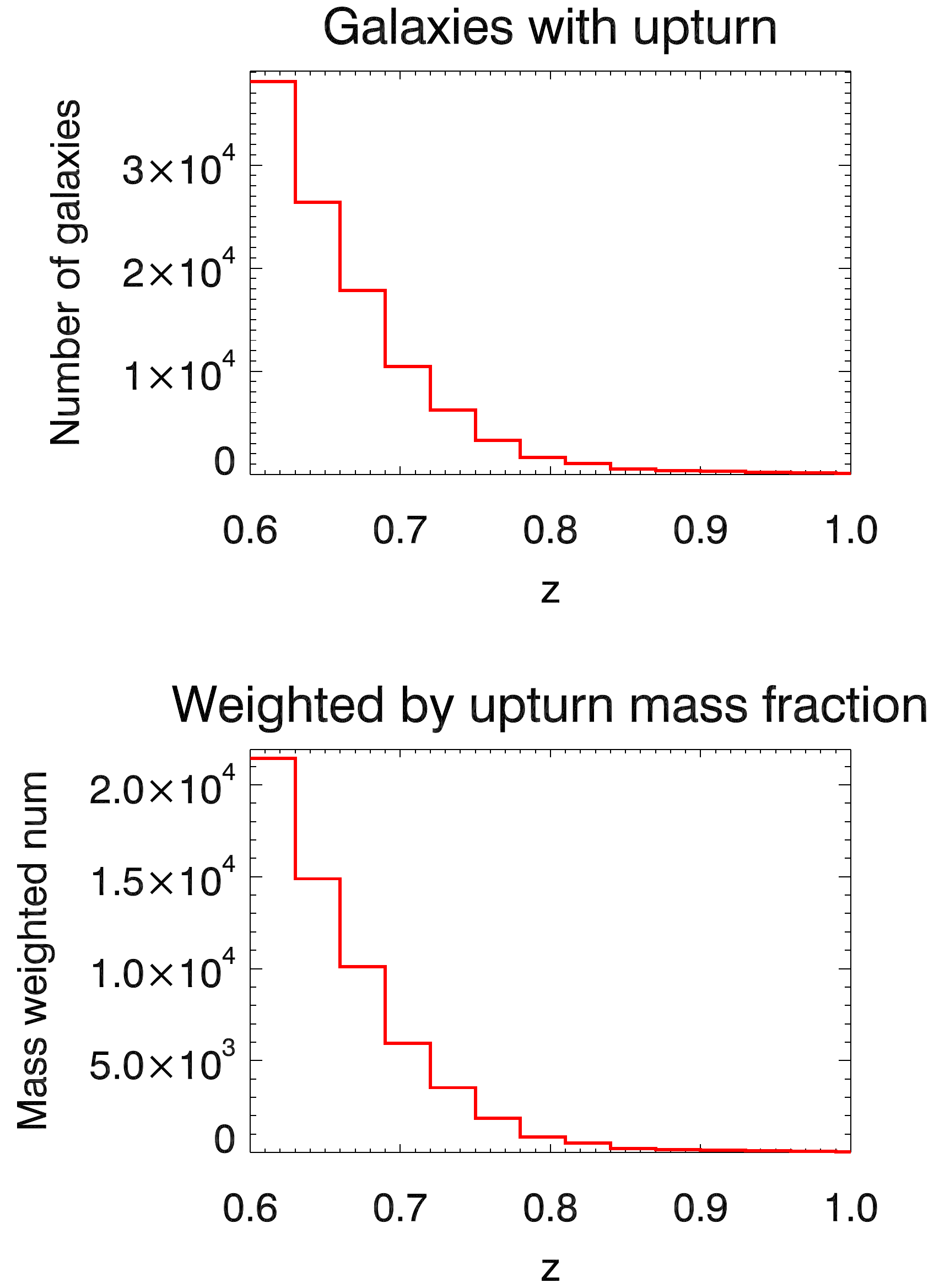}
\caption{\textit{Top panel:} The redshift evolution of the absolute number of galaxies that select an SSP with an old UV-bright population as the best fit solution. \textit{Bottom panel:} The number of galaxies selecting an old UV-bright population weighted by the mass fraction contributing to that population. For details of the mass fractions, see Section \ref{subsec:mass_frac}.}
\label{fig:age_frac_num}
\end{figure}

The distribution of mass fractions shows a slightly different trend when including all upturn contributions, shown by the solid black line, due to the increase in lower mass fractions creating a peak at 20-30\% before decreasing with increasing mass fraction.

\

An investigation into the difference between the properties of galaxies in our upturn confirmed sample and the extra upturn contributions found in {\sc firefly} was undertaken and the results can be seen in Figure \ref{fig:upturn_extra}. On average the galaxies in our upturn confirmed sample are more massive and redder due to the galaxies being older and more likely to have developed a significant upturn. The extra UV flux produced by the UV upturn does not affect the $g - i$ colour of the galaxies as this increase in flux lies well below what the bluest band used for this colour calculation observes.

The redshift evolution of the mass fraction contributing to the upturn for both the upturn confirmed galaxies and the additional galaxies found to have contributions from {\sc firefly} can be seen in Figure \ref{fig:upturn_extra_z}. In both samples there is a trend towards higher mass fractions at lower redshifts. This supports our theory that the upturn starts to occur at $z > 1$ with the strength of the upturn increasing for more evolved galaxies at lower redshifts.

From this analysis it would seem that $\sim$8\% of our working sample do not feature any contribution from the UV upturn. These galaxies show little difference in average redshift $< z >$ and stellar mass $<$ Log(Mass) $>$ when compared to those that select the favoured upturn models (T10 and T25\_2Z) but are significantly redder in the observed-frame $g - i$ colour, obviously. Although this 'odd-one-out' population is a low fraction of the total sample, it deserves further exploration in the future.
\

\subsection{Redshift Evolution of UV upturn}
\label{subsec:z_evo}

Figure \ref{fig:age_frac} shows how the percentage of galaxies preferentially selecting an upturn SSP evolves with redshift. The percentage of galaxies within each redshift bin determined to have an old UV age with an old UV-bright component decreases by $\sim$ 15\%, from $\sim$ 40\% down to $\sim$ 25\%, towards higher redshifts, hinting towards the onset of the UV upturn occurring at $z \approx 1$.

A mass limit was applied to account for any bias in the sample due to observing only the brightest, most massive galaxies at higher $z$. Figure \ref{fig:age_frac_mass} shows the redshift evolution of the percentage of upturn galaxies for galaxies with Log(Mass) $\ge$ 11. A decrease of $\sim$ 18\% is observed in this mass limited sample showing that our result in Figure \ref{fig:age_frac} is not biased by the observational limits of the working sample.

Figure \ref{fig:age_frac_num} shows the absolute number of galaxies that select an upturn SSP with an old UV-bright component over any other in the top panel with the number weighted by the contributing mass fraction shown in the panel below (see Section \ref{subsec:mass_frac} for more details on the mass fractions and their calculation). Both distributions show a steep drop off towards higher $z$ due to the redshift distribution of BOSS galaxies.

\setcounter{figure}{24}
\begin{figure*}
\includegraphics[width=0.95\textwidth]{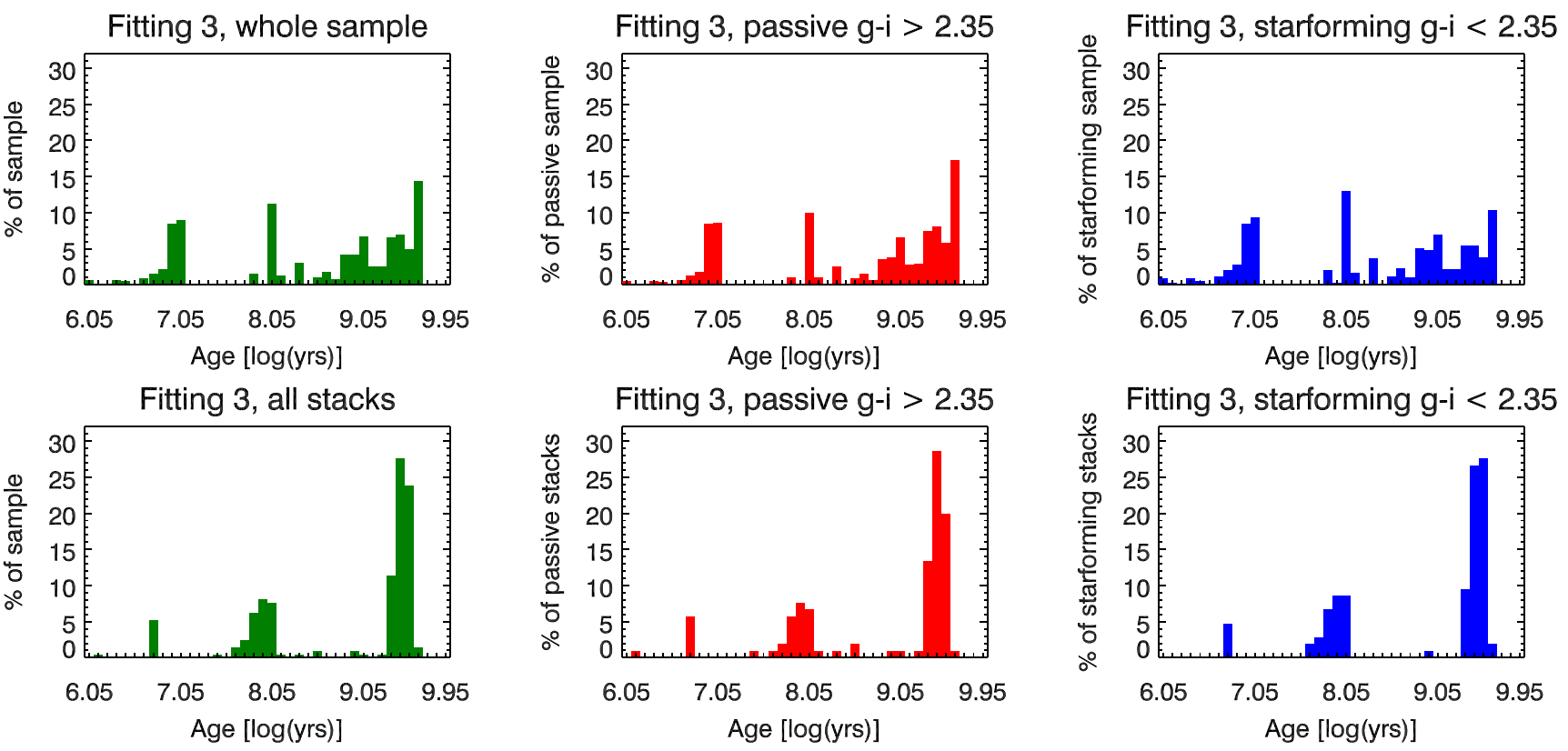}
\caption{Range of UV ages calculated from the fitting of 3 indices shown to be non-degenerate between young and old populations; Mg I, Fe I, and BL3096. Fitting models of upturns found in local galaxies. \textit{LH} -  entire galaxy sample. \textit{M} - galaxies with $g - i > 2.35$ classified as passive ($\sim 51\%$ of entire sample). \textit{RH} - galaxies with $g - i < 2.35$ classified as star forming ($\sim 49\%$ of entire sample) The relevant stacks for each sample are shown in the lower panel.}
\label{fig:ages_local}
\end{figure*}

\section{Searching for high-z analogues of local upturn galaxies}
\label{sec:local}

In order to see if we can find analogues of local upturn galaxies at high $z$ we fit our BOSS galaxy sample using a set of three models found in \citet{Maraston:2000} to fit the upturns found in NGC4472, NGC4649, and the bulge of M31. For more details on the models see Table 1, MT00.

Using the same analysis codes and method of fitting the 3 non-degenerate indices (Mg I, Fe I, and BL3096) we obtain the distribution of UV ages seen in Figure \ref{fig:ages_local}, for both individual galaxies and stacks, using the same colour cut the differentiate between passive and star forming galaxies.

There is less of a peak towards 5 Gyrs in the individual spectra, as seen previously in Figure \ref{fig:hist_cut}, with more of a spread between the ages featured in the upturn models (from 1 Gyr onwards). The spread in ages is also seen in the stacks, with less defined peaks at both old and intermediate ages.

Figure \ref{fig:chi_local} shows the $\chi^{2}$ values found for each of the best fit SSP ages found in Figure \ref{fig:ages_local}. This shows a similar distribution to those found from fitting the original suite of theoretical models. The individual galaxies peak around 3 with the stacks falling around or below 1, with the average values being 9.05 and 1.31 respectively. This is only a slight decrease from the values found fitting the original theoretical models (8.92 and 0.95).

\setcounter{figure}{25}
\begin{figure}
\includegraphics[width=0.5\textwidth]{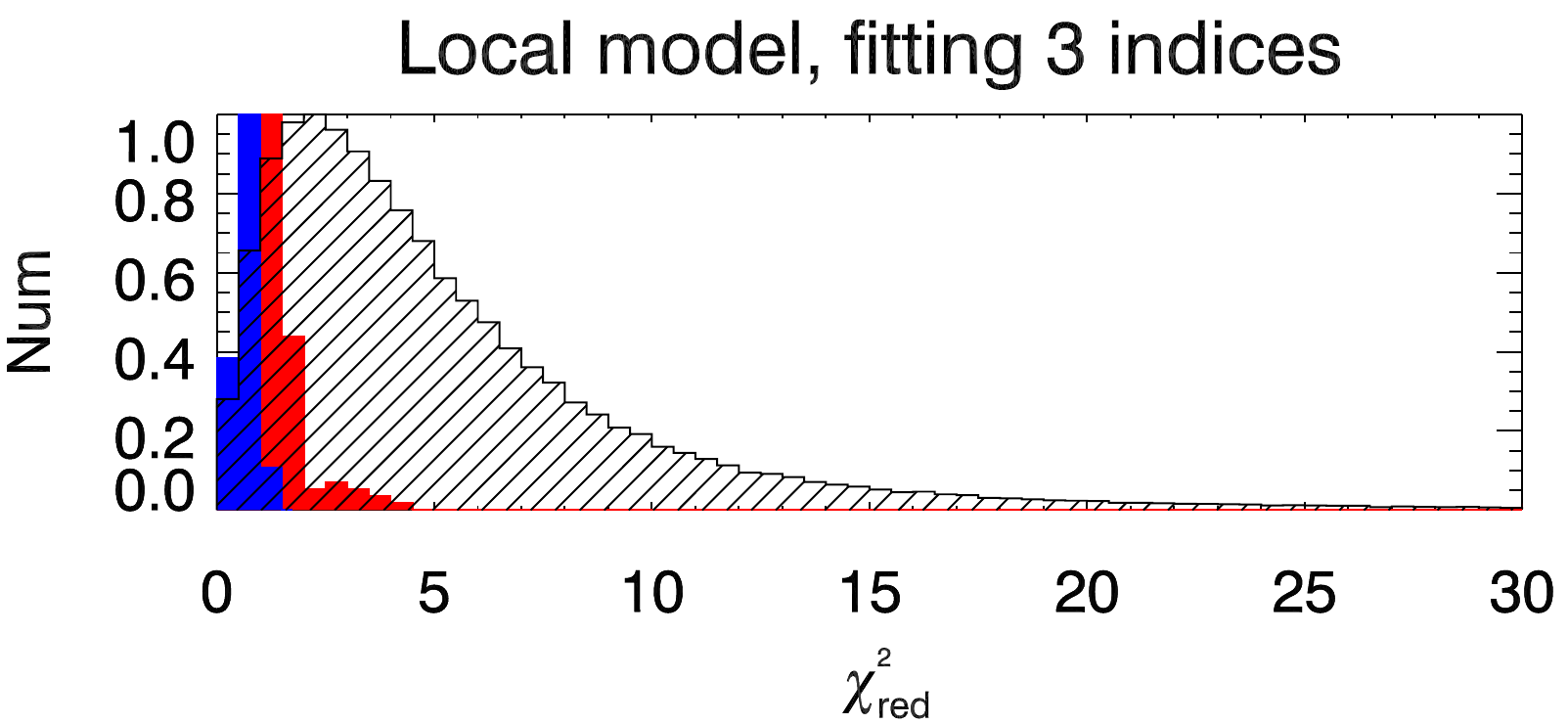}
\caption{Distribution of reduced $\chi^2$ values for the best fits found fitting local upturn models. Fitting the 3 non-degenerate indices (Mg I, Fe I, and BL3096) with the individual spectra shown by the black hashed histograms and the young and old UV age stacks by the solid blue and red histograms respectively. Each histogram has been normalised to 1.}
\label{fig:chi_local}
\end{figure}

Table \ref{tab:local_pref} shows the percentage of the working sample selecting each model, with Table \ref{tab:local_props} giving the average properties of the galaxies.

We find that 39.1\% of our working sample select models of local upturns over young star forming populations. Of these local models, our working sample prefers those of NGC4649, the strongest upturn model in this suite, and NGC4472 with 18.6\% and 16.8\% of our sample selecting each respectively. Both are models of massive elliptical galaxies similar to those selected by BOSS. It is not unexpected that our working sample favours these models over the bulge of M31. In terms of properties, again we find that galaxies selecting young ages have, on average, bluer $g - i$ colours and are less massive.

\setcounter{table}{4}
\begin{table}
\caption{A breakdown of the number of galaxies selecting different local UV models.}
\begin{center}
\begin{tabular}{ | c | c | c | }
Model & Num  & \% of entire sample \\ \hline
\hline
NGC4472 & 46,045 & 16.8 \\
M31 & 10,146 & 3.7 \\
NGC4649 & 51,276 & 18.6 \\
young & 167,194 & 60.9 \\
\label{tab:local_pref}
\end{tabular}
\end{center}
\end{table}

\setcounter{table}{5}
\begin{table}
\caption{Average properties of galaxies selecting different local UV models.}
\begin{center}
\begin{tabular}{ | c | c | c | c | }
Model & $< z >$  & $<$ Log(Mass) $>$ & $< g - i >$ \\ \hline
\hline
NGC4472 & 0.658 & 11.65 & 2.71 \\
M31 & 0.659 & 11.64 & 2.66 \\
NGC4649& 0.660 & 11.62 & 2.63 \\
young & 0.670 & 11.48 & 2.39 \\
\label{tab:local_props}
\end{tabular}
\end{center}
\end{table}

\section{Summary \& Conclusions}
\label{sec:conc}

We have shown results from our application of the \citet{Maraston:2009} modelling of UV absorption line indices to massive galaxies at $z \ge 0.6$. These models include all UV indices defined after the IUE mission (\citet{Fanelli:1992}), and incorporate both a fully theoretical as well as an empirical description of the index strength as a function of stellar parameters. Most indices are robust against dust effects, which make them ideal (in combination) to derive UV ages.

The main scope of this work was to explore models including UV emission from old stars in order to seek for the signature of a UV upturn at high redshift. For this purpose we have extended the M09 models to include a UV upturn component, varying the temperature and energetics of the hot old stars. 

In addition, this paper is also the first application of the modelling of UV spectroscopic indices to quiescent galaxies. 

As observational data we use a large ($\sim 275,000$) sample of galaxies from the Sloan Digital Sky Survey (SDSS) - III Baryon Oscillation Spectroscopic Survey (BOSS, \citet{Dawson:2013}, Data Release 12) as they are ideal for detecting the UV upturn at high-redshift, should it exist. BOSS galaxies are massive (M$>10^{11} \textnormal{M}_{\odot}$), red and passive in large majority ($\sim 80\%$), and distributed to $z \sim$ 1, which, given the wavelength coverage of the BOSS spectrograph (3600 - 10,400\AA) allows us to access mid-UV spectral indices above $z$ = 0.6. The BOSS galaxies are therefore the most likely progenitors of present day massive galaxies which are known to display the UV upturn. 

Early results are promising. First of all, most indices are well resolved in spite of the low S/N ratio ($\sim 5$ in the optical region) of the spectra so that sensible ages and metallicities can be derived. In order to augment the meaningfulness of our quantitative analysis, we perform spectral stacking exploring a wide range of stacking combinations with physical parameters. The stacks have a very high S/N, which allows us to resolve the mid-UV spectral indices such as Fe I, Bl3096, etc. very well. There is a qualitative agreement between the UV age, derived by fitting the absorption indices, and that found from full broadband SED fitting from $u$ to $z$ (previously published as Portsmouth galaxy product, \citet{Maraston:2013boss}). The UV ages are younger than those from full SED fitting and we quantify the offset. Using mock galaxies, we also quantify which age we actually derive when only fitting the UV side of the spectrum.

As one of our main results, we find a set of 3 indices which break the degeneracy between young and old UV ages; Mg I, Fe I, and a blend of several lines, named BL3096.  This is important as the UV upturn can be misinterpreted as residual star formation, with evident implications for our understanding of galaxy formation and evolution processes.

We then use these indices to gain insights into the physical properties of massive galaxies at $z\sim0.6$, which we summarise here.
Calculating the UV ages from these 3 non-degenerate indices alone leads to $\sim48\%$ of our working galaxy sample selecting a model with an upturn component. Forcing the galaxies selecting old UV ages to choose between a model with a contribution from an old hot population and one without results in $\sim78\%$ of the old-age sample selecting the model with old UV-bright stellar populations. This equates to $\sim37\%$ of our entire working sample of BOSS galaxies. 

The lower temperature upturn models, T10 and T25\_2Z, with T$_{eff}\sim$ 10,000 and 25,000 K, and solar or twice solar metallicity, are preferred with 12.1\% and 13.7\% of the working sample selecting ages from each respectively. Interestingly, these models have a fuel consumption which is higher (a factor 3) than what is needed to model strong upturn galaxies locally.\footnote{Although we should note that \citet{Maraston:2000} did not include a temperature as low as 10,000~K in their modelling. Obviously a lower temperature requires a higher fuel consumption.} As local galaxies are several billion years older (12 Gyr vs 3 Gyr), our finding may hint towards an evolutionary effect of the energetics in upturn emitters. The typical temperature instead is consistent with the one appropriate to local galaxies. Moreover, models with a well localised temperature for the upturn are favoured over a blue model with a temperature dispersion (e.g. an extended blue horizontal branch, as found for the low-mass elliptical M32, \citet{Brown:2000b, Brown:2008}). The models with higher fuel values are favoured over those with lower.

An investigation into composite populations has shown that the SSP age calculated from $\chi^2$ minimisation is a good approximation for the UV age of the most dominant UV-bright population in terms of spectral indices. A comparison of age probability distributions from full spectral fitting and fitting Mg I, Fe I, and BL3096, both using a full spectral fitting code ({\sc firefly}), show the existence of composite UV populations in individual galaxies with the percentage of galaxies in our sample with an upturn contribution increasing from $\sim37\%$ to $\sim92\%$. Due to systematics in the data we have shown fitting the spectral energy distribution between 2200 - 3200\AA \ to be unreliable. Using composite model fitting we can also determine the fraction of galaxy mass hosting UV upturn populations. This fraction peaks at 20-30\%.

Upturn confirmed galaxies are on average redder and more massive. Galaxies with higher mass fractions contributing to the upturn are also found to be redder and more massive. These trends are in qualitative agreement with the trend of the UV upturn locally. 

We further test for high-$z$ analogues of local galaxies by fitting the same suite of models known to best fit the UV upturn seen in the local massive elliptical galaxies NGC4472, NGC4649, and the bulge of M31. Not surprisingly, we find these models to fit our sample equally well as the theoretical suite produced for this paper.

However we find many more galaxies selecting an upturn model than what is deduced studying local samples using photometric indicators (e.g. the 5\% from \citet{Yi:2005}). We cannot single out the exact reason for this discrepancy, but we can envisage a number of factors contributing to a different result. First, we use a sample of really massive galaxies, whereas locally one has access to much fewer truly massive objects simply due to volume effects and sample selection (\citet{Bureau:2011}). Then, we use spectral indices whereas local studies are mostly based on broad-band photometry. Also, our analysis includes a large variation in the models. Using mock galaxies, for example, we demonstrate that when a galaxy hosting UV light is analysed solely with classical, non upturn models, the derived age is young ($\sim$ 2 Gyr). If the same galaxy is also fitted with upturn models, the derived age is old and coincides  with the mock age of input. This is to say that possibly galaxies hosting a UV upturn in small proportions are assigned a younger age when they are analysed with a restricted suite of models. This may explain the situation of \citet{Yi:2005, Yi:2011} who were not able to classify a large fraction of their UV-detected galaxies. A proper addressing of the comparison between the distant and local universe might have to wait for future spectroscopic facilities in the ultraviolet.

Finally, we determine the redshift evolution of the upturn population. This expands on the pioneeristic work of \citet{Brown:1998, Brown:2000, Brown:2003}, who follow the evolution of the upturn using far-UV images of a few (4 - 8) elliptical galaxies at $z$ = 0.375, 0.55, and 0.33 respectively. They find little evolution of the strength of the upturn over this redshift range but with galaxies at $z$ = 0.33 and $z$ = 0.55 having upturns weaker than those found in local galaxies.
Looking at higher redshift ($z > 0.6$) with a much larger sample of galaxies and using spectroscopy, we find the percentage of galaxies preferentially selecting an upturn over all others to evolve with redshift. The percentage decreases from $\sim$40\% at $z \sim$ 0.6 to $\sim$ 25\% at $z \sim$ 1. This hints towards to onset of such populations at $z > 1$. The increase of galaxies hosting an upturn population towards our epoch ($z\sim 0$) is consistent with stellar evolution, as the typical stellar mass in galaxies decreases with decreasing redshift.

\section*{Acknowledgements}
We acknowledge discussion with Tom Brown.
Funding for SDSS-III has been provided by the Alfred P. Sloan Foundation, the Participating Institutions, the National Science Foundation, and the U.S. Department of Energy Office of Science. The SDSS-III web site is http://www.sdss3.org/.

SDSS-III is managed by the Astrophysical Research Consortium for the Participating Institutions of the SDSS-III Collaboration including the University of Arizona, the Brazilian Participation Group, Brookhaven National Laboratory, Carnegie Mellon University, University of Florida, the French Participation Group, the German Participation Group, Harvard University, the Instituto de Astrofisica de Canarias, the Michigan State/Notre Dame/JINA Participation Group, Johns Hopkins University, Lawrence Berkeley National Laboratory, Max Planck Institute for Astrophysics, Max Planck Institute for Extraterrestrial Physics, New Mexico State University, New York University, Ohio State University, Pennsylvania State University, University of Portsmouth, Princeton University, the Spanish Participation Group, University of Tokyo, University of Utah, Vanderbilt University, University of Virginia, University of Washington, and Yale University. 

\appendix


\section{Testing the Effect of Dust Using Mock Galaxies}
\label{subsec:dust}

Reddening due to interstellar dust is pronounced in the UV region, which could effect several indices. To investigate this effect on the calculation of the UV age we tested our age derivation based on indices using mock galaxies with well-defined input properties calculated with the \citet{Maraston:2005} software. We used composite stellar populations (CSPs) with a nearly constant star formation rate (\textbf{star formation timescale} $\tau$ = 10 Gyrs), solar metallicity and an age of 0.1 Gyrs. To maximise any effect from dust we selected two mock galaxies, one with no dust and one with a dust attenuation of  $a=3$A$_{v}$ from a Calzetti law.

We calculated the EWs of the mid-UV indices from the mock galaxy spectra and derived the UV ages from fitting all 7 indices simultaneously and the 3 indices found to be non-degenerate between young and old UV-bright populations (Mg I, Fe I, and BL3096) as well as each index separately so as to highlight any specific indices affected. Figure \ref{fig:dust} shows the resulting ages for the theoretical models.

One index, Fe I, shows significant offset from the one-to-one relationship shown in black, with the age decreasing by log(age [yrs]) = 1.88 with the inclusion of dust. Fe II 2402, BL2538, and BL3096 show slight offsets with the first two increasing in age with the addition of dust and the last decreasing.

Despite these affected indices the UV ages derived when fitting a combination of indices, shown by the black plus and square, there is negligible effect on the calculated age due to the majority of indices showing little to no dependence on dust.  From this we conclude that when several indices are used together there is negligible effect from dust on the derived galaxy age. However in the case of individual indices we must be careful with those shown to change significantly.

\setcounter{figure}{0}
\begin{figure}
\includegraphics[width=0.45\textwidth]{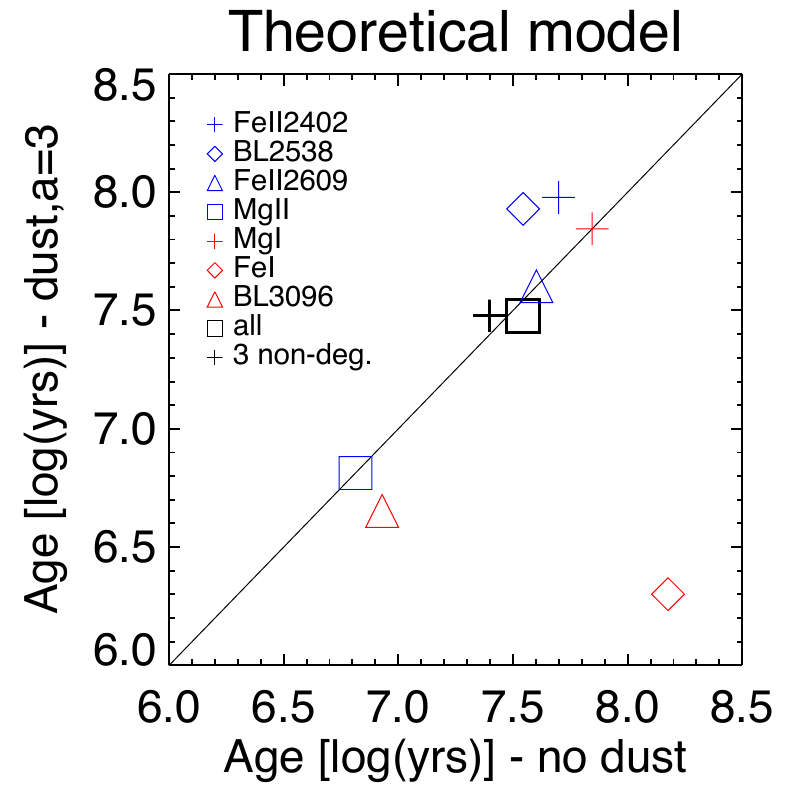}
\caption{The effect of dust on the calculation of the UV age. The age derived from each individual index is shown for two mock galaxies, one with no dust, and one with high dust attenuation ($a=3$A$_{v}$). The indices used for each point are shown in the legend.}
\label{fig:dust}
\end{figure}

\bibliographystyle{mnras}

\providecommand{\aj}{Astron. J. }\providecommand{\apj}{Astrophys. J.
  }\providecommand{\apjs}{Astrophys. J.
  }\providecommand{\mnras}{MNRAS}\providecommand{\aap}{Astron. Astrophys.
  }\providecommand{\aaps}{Astron. Astrophys. }\providecommand{\fcp}{Fundam.
  Cosm. Phys. }\providecommand{\araa}{Annu. Rev. Astron.
  Astrophys.}\providecommand{\pasp}{PASP}\providecommand{\apss}{Astrophys.
  Space Sci.}

\clearpage  

\end{document}